\documentclass[12pt]{article}

\usepackage[height=8.85in,width=6.45in]{geometry}

\usepackage[utf8]{inputenc}
\usepackage[T1]{fontenc}
\usepackage{amsmath}
\usepackage{amssymb}
\usepackage{mathtools}
\numberwithin{equation}{section}
\allowdisplaybreaks

\usepackage{slashed}
\usepackage{braket}
\usepackage[usenames,dvipsnames,svgnames,table]{xcolor}
\usepackage[colorlinks,citecolor=DarkGreen,linkcolor=FireBrick,urlcolor=FireBrick,linktocpage]{hyperref}
\usepackage{cite}
\usepackage{graphicx}
\usepackage{times}
\usepackage[scaled]{couriers}

\usepackage{bm}
\usepackage{subfig}

\usepackage{tikz,pgf}
\usepackage{tikz-cd}
\usetikzlibrary{shapes}
\usetikzlibrary{calc}
\usetikzlibrary{decorations.pathmorphing}
\usetikzlibrary{decorations.pathreplacing,shapes.misc}
\usetikzlibrary{positioning}
\usetikzlibrary{decorations.markings}
\tikzset{->-/.style={decoration={
  markings,
  mark=at position .5 with {\arrow{>}}},postaction={decorate}}}
\tikzset{-<-/.style={decoration={
  markings,
  mark=at position .5 with {\arrow{<}}},postaction={decorate}}}

\usepackage{xcolor}
  \definecolor{rblue}{RGB}{81, 49, 193}
  \definecolor{rorange}{RGB}{255, 147, 40}
  \definecolor{rgreen}{RGB}{176, 233, 0}

\renewcommand{\tilde}{\widetilde}

\usepackage{mdframed}

\begin{document}

\begin{center}

{\large \bfseries Commuting projector models for (3+1)d topological superconductors via string net of (1+1)d topological superconductors}

\bigskip
\bigskip
\bigskip

Ryohei Kobayashi
\bigskip
\bigskip
\bigskip

\begin{tabular}{ll}
 Institute for Solid State Physics, \\
University of Tokyo, Kashiwa, Chiba 277-8583, Japan\\

\end{tabular}

\vskip 1cm

\end{center}

\noindent 
We discuss a way to construct a commuting projector Hamiltonian model for a (3+1)d topological superconductor in class DIII.
The wave function is given by a sort of string net of the Kitaev wire, decorated on the time reversal (T) domain wall.
Our Hamiltonian is provided on a generic 3d manifold equipped with a discrete form of the spin structure.
We will see how the 3d spin structure induces a 2d spin structure (called a “Kasteleyn” direction on a 2d lattice) on T domain walls, which makes possible to define fluctuating Kitaev wires on them.
Upon breaking the T symmetry in our model, we find the unbroken remnant of the symmetry which is defined on the time reversal domain wall.
The domain wall supports the 2d non-trivial SPT protected by the unbroken symmetry, which allows us to determine the SPT classification of our model,
based on the recent QFT argument by Hason, Komargodski, and Thorngren.

\setcounter{tocdepth}{2}
\tableofcontents

\section{Introduction}

The notion of fermionic topological phases of matter has attracted great interest, since fermionic systems host novel phases that have no counterpart in bosonic systems~\cite{Gu:2012ib, Wang2017Interacting, Witten2016Fermion, Metlitski2014, Cheng2018Classification, Gu2014Lattice, Guo:2018vij, Gaiotto:2015zta}. Of particular interest are invertible topological phases, which feature a unique ground state on a closed spatial manifold. 
In the presence of global symmetries, invertible topological phases are sometimes called Symmetry Protected Topological (SPT) phases. Fermionic SPT phases are thought to be described by spin/pin invertible Topological Quantum Field Theory (TQFT) at long distances~\cite{Gaiotto:2015zta, Bhardwaj2017Statesum, Bhardwaj2017, Turzillo2018, Kobayashi2019pin}, which is classified by the spin/pin cobordism group up to symmetric deformation~\cite{Kapustin:2014dxa, Freed:2016rqq, Yonekura:2018ufj}.

A rather well-understood class of $(d+1)$-dimensional fermionic SPT phases are classified by group supercohomology~\cite{Gu:2012ib}. While covering a large class of SPT phases, the classification leaves out ``beyond supercohomology'' phases, whose classification was developed in~\cite{wanggu1703, wanggu1811}.
The simplest beyond supercohomology phase is the (1+1)d topological superconductor (Kitaev wire) in (1+1)d~\cite{Kitaev00unpaired}. In the absence of global symmetries except for fermion parity, the Kitaev wire generates the $\mathbb{Z}_2$ classification of SPT phases. If we take a time reversal symmetry with $T^2=1$ into account, it instead generates the $\mathbb{Z}_8$ classification~\cite{FidkowskiKitaev2011}.

In (2+1)d, there is a way to provide an exactly solvable model for a beyond supercohomology SPT protected by the unitary $\mathbb{Z}_2$ symmetry, on a graph whose edges are directed in a specific way~\cite{Tarantino}. The wave function for this phase is described as a sort of string net of the Kitaev wire. 
Concretely, the phase is given by first decorating the Kitaev wire on the $\mathbb{Z}_2$ domain wall, and then fluctuating the domains to respect the $\mathbb{Z}_2$ symmetry. In order to conserve fermion parity under fluctuation of the Kitaev wire, one requires a specific choice of directions on edges called ``Kasteleyn direction'', which is understood as a discrete form of the spin structure on a spatial manifold~\cite{David06dimer}.

\begin{figure}[htb]
\centering
\includegraphics{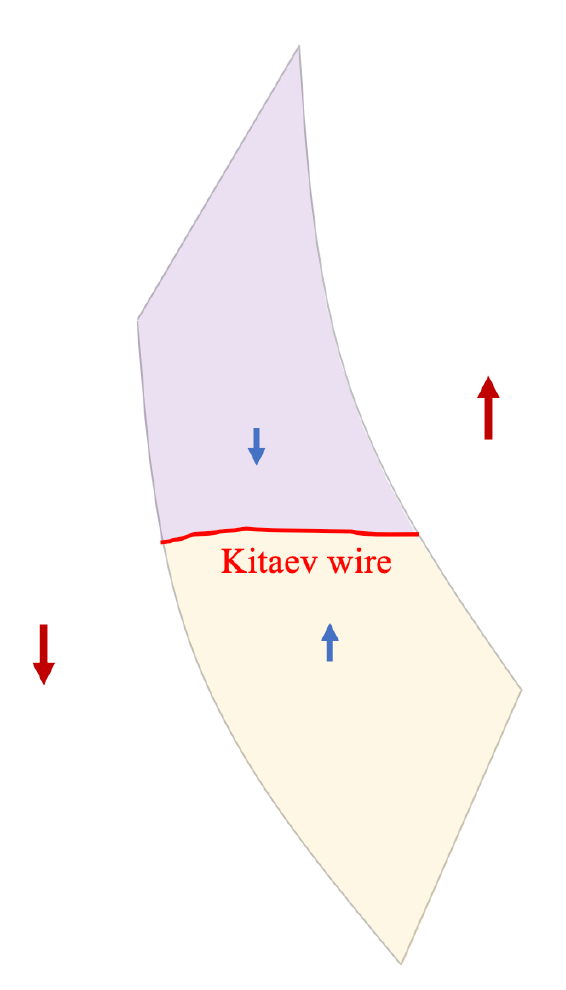}
\caption{A schematic illustration for our wave function in 3d. On a 2d $T$ domain wall, we have a 2d wave function of fluctuating Kitaev wires. The domains are characterized by qubits (red and blue arrows) charged by the time reversal symmetry.}
\label{fig:schematic}
\end{figure}

In this note, as a generalization of the prescription in (2+1)d~\cite{Tarantino}, we will describe (3+1)d topological superconductors beyond supercohomology, in terms of the Kitaev wire decorations.
More concretely, we focus on the well-known $\mathbb{Z}_{16}$ classification of (3+1)d topological superconductor protected by the time reversal symmetry with $T^2=(-1)^F$ (class DIII)~\cite{FidkowskiChenVishwanath2014}, and provide a way to generate a $\mathbb{Z}_8$ subclass of the $\mathbb{Z}_{16}$ classification based on the string net of the Kitaev wire. 
Our model is also understood as a version of decorated domain wall construction~\cite{Chen13decorated, Ware16dimer, Nat18full}, where the $T$ domain wall ferries a 2d wave function of the fluctuating Kitaev wires, see Fig.~\ref{fig:schematic}. 
By deliberately assigning the directions on edges of the 3d graph, we always have a 2d graph on the $T$ domain wall whose edges are completely Kasteleyn directed, allowing us to fluctuate Kitaev wires on the wall in a fashion respecting fermion parity.

We will see that the Kasteleyn property of the arbitrary 2d domain wall is made possible by a choice of a 3d discrete form of the spin structure on a spatial manifold.
We explicitly provide a commuting projector Hamiltonian which produces the above string net picture at the ground state, defined on any closed oriented 3d spin manifolds equipped with a triangulation.

To figure out what phase our model is in, we make contact with the recent QFT argument which determines the SPT classification from that for the domain wall~\cite{HKT2019CPT, COSY2019decorated}. 
If we prepare an underlying TQFT for our model with the time reversal symmetry $T^2=(-1)^F$, the TQFT is defined on a spacetime equipped with a pin$^{+}$ structure.
There is an induced spacetime structure on a (2+1)d domain wall~\cite{Kapustin:2014dxa}, which implies the existence of the symmetry induced on the wall. In this case, the induced structure corresponds to unitary $\mathbb{Z}_2$ symmetry.

Once the configuration of the (2+1)d domain wall is fixed by spontaneous breaking of the $T$ symmetry, we observe the effective theory for the domain wall enjoying the induced $\mathbb{Z}_2$ symmetry. Especially, the domain wall in general supports a (2+1)d SPT phase protected by the induced symmetry, if we start from (3+1)d $T$-SPT phase.
Interestingly, the classification of the (2+1)d $\mathbb{Z}_2$-SPT on the domain wall completely determines that of the (3+1)d $T$-SPT~\cite{HKT2019CPT}.

In our commuting projector model, we claim that the domain wall supports a nontrivial SPT based on the induced $\mathbb{Z}_2$ symmetry, after fixing the configuration of the wall by breaking the $T$ symmetry. 
The classification of the (2+1)d fermionic $\mathbb{Z}_2$-SPT phase is given by cobordism group $\Omega_{\mathrm{spin}}^3(B\mathbb{Z}_2)=\mathbb{Z}_8\times\mathbb{Z}$. The domain wall is effectively described by the commuting projector model given in~\cite{Tarantino} for the $\mathbb{Z}_8$ root phase with the trivial $\mathbb{Z}$ index, generating the $\mathbb{Z}_8$ classification.
Then, the classification of the (3+1)d $T$-SPT phase is totally encoded in the SPT on the domain wall, which allows us to find that our model realizes the $\mathbb{Z}_8$ root phase of the $\mathbb{Z}_{16}$ classification.

We do not produce the generator of the full $\mathbb{Z}_{16}$ classification in the present paper, which requires the (2+1)d domain wall to support the SPT phase with the nonzero, especially odd $\mathbb{Z}$ index. This corresponds to stacking of $p+ip$ superconductors, which is unlikely to be realized by a commuting projector Hamiltonian~\cite{Spodyneiko}.

As an application of our construction, we also provide a (3+1)d $\mathbb{Z}_4^F$ symmetric model described as the string net of the Kitaev wire, where we also find an induced symmetry on the domain wall.

This note is organized as follows. In Sec.~\ref{sec:tarantino}, we review the construction of the (2+1)d $\mathbb{Z}_2$-SPT phase in terms of the domain wall decoration of the Kitaev wire. 
In Sec.~\ref{sec:3dt}, we propose a (3+1)d $T$-SPT model defined on a 3d closed oriented manifold with a spin structure. This model is regarded as decorating the (2+1)d model in Sec.~\ref{sec:tarantino} on $T$ domain walls.
In Sec.~\ref{sec:3dz4f}, we discuss the generalization to the (3+1)d $\mathbb{Z}_4^F$ symmetric phase. In Sec.~\ref{sec:wallspt}, we determine the SPT classification of our model by studying the domain wall symmetry.

\section{Review: lattice model of (2+1)d $\mathbb{Z}_2$ SPT}
\label{sec:tarantino}
In this section, we first recall the construction of the (2+1)d $\mathbb{Z}_2$ SPT phase by Tarantino and Fidkowski~\cite{Tarantino}. Though the model was originally build on a honeycomb lattice in~\cite{Tarantino}, we refer to the model on any 2d oriented manifold equipped with a triangulation, which was obtained in~\cite{wanggu1703}. 

\begin{figure}[htb]
\centering
\includegraphics{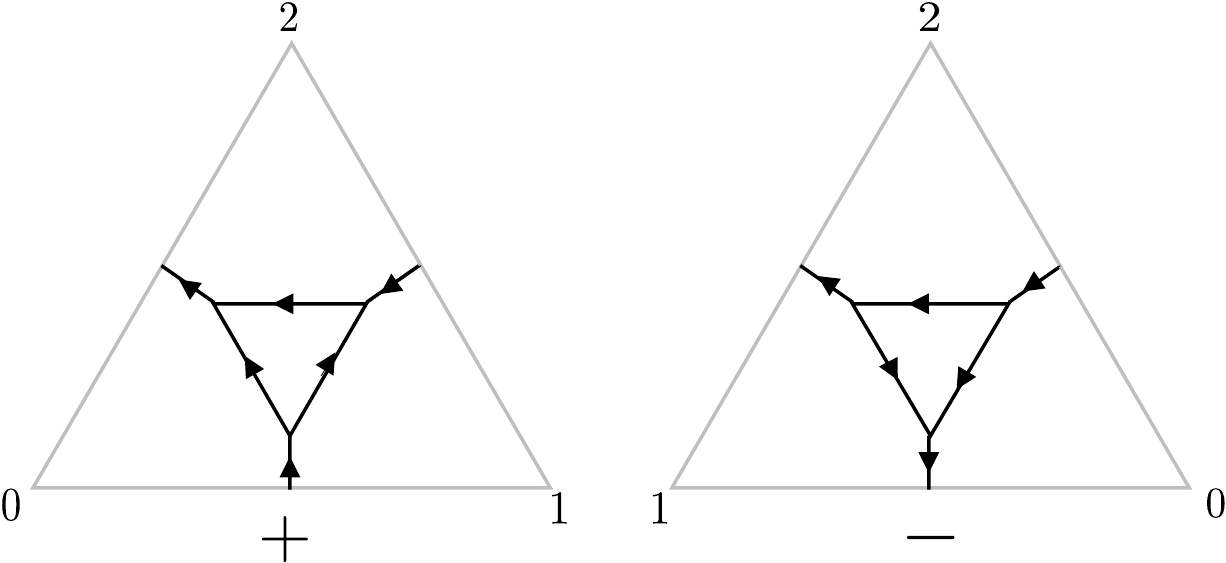}
\caption{Directions on edges of $\Gamma$.}
\label{fig:2ddirection}
\end{figure}

We consider a trivalent directed graph $\Gamma$ on a 2d oriented spin manifold $M$ given as follows. 
We consider a triangulated $M$ with a branching structure. The simplical complex for this triangulation is denoted as $\mathcal{T}$. We have local ordering on each 2-simplex of $\mathcal{T}$ according to the branching structure.
Each 2-simplex can then be either a $+$ simplex or a $-$ simplex, depending on whether the ordering agrees with the orientation or not.
Then, the trivalent graph $\Gamma$ is obtained by filling each 2-simplex of $\mathcal{T}$ with a pattern described in Fig.~\ref{fig:2ddirection}. The edges of $\Gamma$ are Kasteleyn directed, and they are assigned in the following steps~\cite{wanggu1703};
\begin{enumerate}
    \item We start with directing edges of the graph $\Gamma$, as described in Fig.~\ref{fig:2ddirection}, for $+$ and $-$ simplices of $\mathcal{T}$. At this stage, some faces of $\Gamma$ are not necessarily Kasteleyn.
    \item Each non-triangular face of $\Gamma$ is in 1-1 correspondence to a 0-simplex of $\mathcal{T}$. We denote as $S$ the set of 0-simplices that correspond to non-Kasteleyn faces of $\Gamma$. $S\in Z_0(M, \mathbb{Z}_2)$ is known to represent the dual of the second Stiefel-Whitney class $w_2$~\cite{Goldstein, Thorngrenthesis}. Then, we specify the spin structure of $M$ by a choice of a set of 1-simplices $E\in C_1(M,\mathbb{Z}_2)$ of $\mathcal{T}$, with $\partial E=S$.
    \item We reverse the directions on the edge of $\Gamma$ which cross 1-simplices of $E$, which makes all faces of $\Gamma$ Kasteleyn.
\end{enumerate}

After these steps, we introduced the Kasteleyn direction on edges of $\Gamma$ based on the spin structure of $M$. We refer to triangular faces in $\Gamma$ as triangles. Let $t(v), t(w)$ be triangles in $\Gamma$ that contain vertices $v, w$ respectively. If an edge $\langle vw\rangle$ satisfies $t(v)=t(w)$, we refer to $\langle vw\rangle$ as a long edge. If we have $t(v)\neq t(w)$, we refer to $\langle vw\rangle$ as a short edge.
Then, the degrees of freedom in the model are given as follows:
\begin{itemize}
    \item A qubit located on each face of $\Gamma$ except for triangles, operated by Pauli operators $\tau^x, \tau^y, \tau^z$.
    Dually, we can also think of $\tau$ qubits as living on 0-simplices of $\mathcal{T}$.
    
    \item A complex fermion on each short edge $e$ of $\Gamma$, created and annihilated by $c^{\dagger}_e, c_e$ respectively.
    Dually, we can also think of complex fermions as living on 1-simplices of $\mathcal{T}$.
\end{itemize}
The qubits are charged under the unitary $\mathbb{Z}_2$ symmetry, $U_{\mathbb{Z}_2}: \ket{1}\mapsto\ket{0}, \ket{0}\mapsto\ket{1}$, while the fermions are invariant under the symmetry. 
The wave function for this model is given by decorating the Kitaev wire on the 1d domain wall of qubits. To introduce the Kitaev wire decoration, it is convenient to decompose each complex fermion into a pair of Majorana fermions. 
Let $e = \langle \overrightarrow{vv'}\rangle$ be a short edge oriented from $v$ to $v'$. Then, each complex fermion on $e$ is represented by a pair of Majorana fermions
\begin{align}
    \begin{split}
        a_{v} &= c_{e}^{\dagger}+c_{e}, \\
        b_{v'} &= i(c_{e}^{\dagger}-c_{e}),
    \end{split}
\end{align}
located on $v$ and $v'$, respectively. The wave function is then given by pairing Majorana fermions on vertices of $\Gamma$, according to the dimer covering of the edges of $\Gamma$. 

\begin{figure}[htb]
\centering
\includegraphics{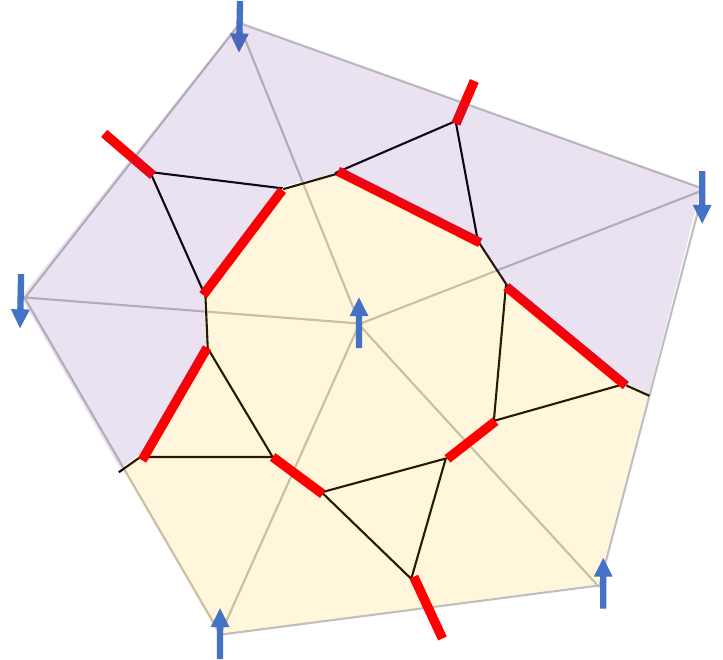}
\caption{Majorana pairings on the 2d graph $\Gamma$.}
\label{fig:dimer2d}
\end{figure}

As a technical detail, we locate a fictitious qubit on each triangle of $\Gamma$, whose $\tau^z$ is fixed according to the majority rule: if the triangle is contained in a 2-simplex $\langle v_0v_1v_2\rangle$, it is $\ket{1}$ or $\ket{0}$ depending on whether the majority of three qubits on $v_1,v_2,v_3$ have $\ket{1}$ or $\ket{0}$.
Then, away from the domain wall of qubits, we pair up Majorana fermions along short edges $\langle \overrightarrow{vw}\rangle$ by a pairing term $i\gamma_v\gamma_w$ ($\gamma$ is $a$ or $b$). On the domain wall, we pair up Majorana fermions along long edges $\langle\overrightarrow{vw}\rangle$ by a pairing term $i\gamma_v\gamma_w$. These pairing rules amount to decorating the Kitaev wire on the 1d domain wall of qubits; see Fig.~\ref{fig:dimer2d}.
The wave function of the model is given by the equal superposition of all possible configurations of qubits, associated with the Majorana pairings discussed above.

\subsection{Kasteleyn direction and conservation of fermion parity}
\label{subsec:kasteleyn}
Before providing the Hamiltonian for the wave function, let us comment on why we need the Kasteleyn direction for the construction.
In short, the Kasteleyn property of $\Gamma$ is required in order to have the $\mathbb{Z}_2^F$ symmetric state.

To see this, let us consider a wave function with a specific pairing of Majorana fermions, which corresponds to a dimer covering $\mathcal{D}_i$ on $\Gamma$. 
By flipping some qubits for this wave function, we finally obtain a different dimer covering $\mathcal{D}_f$. These two dimer coverings are related by sliding a sequence of dimers along a closed path $C$ of $\Gamma$. Concretely, suppose edges $\langle v_1v_2\rangle$, $\langle v_3v_4\rangle$, $\dots\langle v_{2n-1}v_{2n}\rangle$ form dimers in $\mathcal{D}_i$. Then, the dimers are rearranged to $\langle v_2v_3\rangle$, $\langle v_4v_5\rangle$, $\dots\langle v_{2n}v_{1}\rangle$ in $\mathcal{D}_f$. 
We can easily show that the two wave functions for $\mathcal{D}_i$ and $\mathcal{D}_i$ have the same fermion parity, \textit{iff} the path $C$ is Kasteleyn directed.
If we work on the reduced Fock space of these $2n$ Majorana fermions on $C$, the fermion parity for $\mathcal{D}_i$ is
\begin{align}
    i^n\gamma_1\gamma_2\dots\gamma_{2n-1}\gamma_{2n}=s_{1,2}s_{3,4}\dots s_{2n-1,2n},
\end{align}
where $s_{i,j}=1$ if the direction for $\langle v_iv_j\rangle$ is $\langle\overrightarrow{v_iv_j}\rangle$,
and $s_{i,j}=-1$ for the opposite direction.
The fermion parity for $\mathcal{D}_f$ is
\begin{align}
    i^n\gamma_1\gamma_2\dots\gamma_{2n-1}\gamma_{2n}=-i^n\gamma_2\dots\gamma_{2n-1}\gamma_{2n}\gamma_1=-s_{2,3}s_{4,5}\dots s_{2n,1}.
\end{align}
These two expressions are identical iff $C$ is Kasteleyn directed.

\subsection{Hamiltonian}
Now we provide the Hamiltonian for the (2+1)d model. The Hamiltonian consists of two terms,
\begin{align}
    H= H_{\mathrm{decorate}}+H_{\mathrm{fluct}}.
    \label{eq:ham2}
\end{align}

\subsubsection{$H_{\mathrm{decorate}}$}
The term $H_{\mathrm{decorate}}$ pairs up Majorana fermions according to the configuration of qubits, which has the form of
\begin{align}
    H_{\mathrm{decorate}} &=\sum_{\substack{\mathrm{short\ edge}\\ \langle \overrightarrow{vw}\rangle}}i\gamma_v\gamma_w(1-D_{\langle vw\rangle})+\sum_{\substack{\mathrm{long\ edge}\\ \langle \overrightarrow{vw}\rangle}}i\gamma_v\gamma_wD_{\langle vw\rangle}
\end{align}
We define $D_{\langle vw\rangle}:=(1-\tau_f^z\tau_{f'}^z)/2$ where $f$, $f'$ are faces sandwiching the edge $\langle vw\rangle$.
$D_{\langle vw\rangle}$ returns 1 when we decorate the Kitaev wire on $\langle vw\rangle$, otherwise it returns 0.

\subsubsection{$H_{\mathrm{fluct}}$}
$H_{\mathrm{fluct}}$ will be defined to tunnel between the different configurations of qubits, which has the form of
\begin{align}
    H_{\mathrm{fluct}}=\sum_f \tau_f^x X_f,
\end{align}
where $X_f$ denotes an operator which rearranges the dimer configuration of Majorana fermions, associated with the bit flip $\tau_f^x$ at the face $f$,
\begin{align}
    X_f=\sum_{\{t^z\};  R_f}X_f^{\{t^z\}}\Pi_f^{\{t^z\}}P_f^{\{t^z\}}.
    \label{eq:xf2d}
\end{align}
Here, $R_f$ denotes a set of qubits whose eigenvalues of $\tau^z$ determine the pairing rule of Majorana fermions on vertices of $f$. Specifically, we define $R_f$ as the local set of qubits containing those on $f$ and all faces adjacent to $f$. 
Then, the sum in~\eqref{eq:xf2d} is over $2^{|R_f|}$ patterns of $\tau^z$ eigenvalues $\{t^z\}$ in $R_f$.  $P_{f}^{\{t^z\}}$ is a projector for qubits in $R_f$ which stabilizes a given set of eigenvalues $\{t^z_f\}$ of $\{\tau^z_f\}$ in the summand. $\Pi_f^{\{t^z\}}$ is a projector for the fermionic Hilbert space which stabilizes the preferred Majorana pairings by $H_{\mathrm{decorate}}$,
\begin{align}
    P_{f}^{\{t^z\}}=\prod_{\mathrm{qubits}\in R_f}
    \frac{1+t^z_f\tau^z_f}{2},
    \label{eq:pf}
\end{align}
\begin{align}
\begin{split}
    \Pi_{f}^{\{t^z\}}=&\prod_{\mathrm{short}\ \langle \overrightarrow{vw}\rangle\in C}\left(\frac{1+i\gamma_v\gamma_w}{2}(1-D_{\langle vw\rangle})+D_{\langle vw\rangle}\right)\\
    \cdot&\prod_{\mathrm{long}\ \langle \overrightarrow{vw}\rangle\in C}\left((1-D_{\langle vw\rangle})+\frac{1+i\gamma_v\gamma_w}{2}D_{\langle vw\rangle}\right),
    \end{split}
\end{align}
where $C$ denotes a closed path of $\Gamma$ where the rearrangement of Majorana pairings takes place.

Then, $X_f^{\{t^z\}}$ has the effect of moving Majorana pairings along $C$. Following the notations in Sec.~\ref{subsec:kasteleyn}, we can express $X_f$ as
\begin{align}
    X_f^{\{t^z\}}=2^{-\frac{n+1}{2}}(1+is_{2,3}\gamma_2\gamma_3)(1+is_{4,5}\gamma_4\gamma_5)\dots(1+is_{2n,1}\gamma_{2n}\gamma_1),
\end{align}
where $C$ has the length $2n$, and the final dimer configuration after acting the bit flip $\tau^x_f$ has the dimers on $\langle v_2v_3\rangle$, $\langle v_4v_5\rangle$, $\dots\langle v_{2n}v_{1}\rangle$.

\section{Lattice model of (3+1)d time-reversal SPT}
\label{sec:3dt}
We will generate a lattice model for a (3+1)d time-reversal SPT phase with $T^2=(-1)^F$, which is provided by a sort of decorated domain wall construction of the Kitaev wire.
We consider a trivalent graph $\Gamma$ on a 3d oriented spin manifold $M$ given as follows. 

We first endow $M$ with a triangulation. In addition, we take the
barycentric subdivision for the triangulation of $M$. Namely, each 3-simplex in the initial triangulation of $M$ is subdivided into $4!=24$ simplices, whose vertices are barycenters of the subsets of vertices in the $3$-simplex. 
We further assign a local ordering to vertices of the barycentric subdivision, such that a vertex on the barycenter of each $i$-simplex is labeled as ``$i$.''
The obtained simplical complex after taking barycentric subdivision is denoted as $\mathcal{T}$.
Each 3-simplex can then be either a $+$ simplex or a $-$ simplex, depending on whether the ordering agrees with the orientation or not.

\begin{figure}[htb]
\centering
\includegraphics{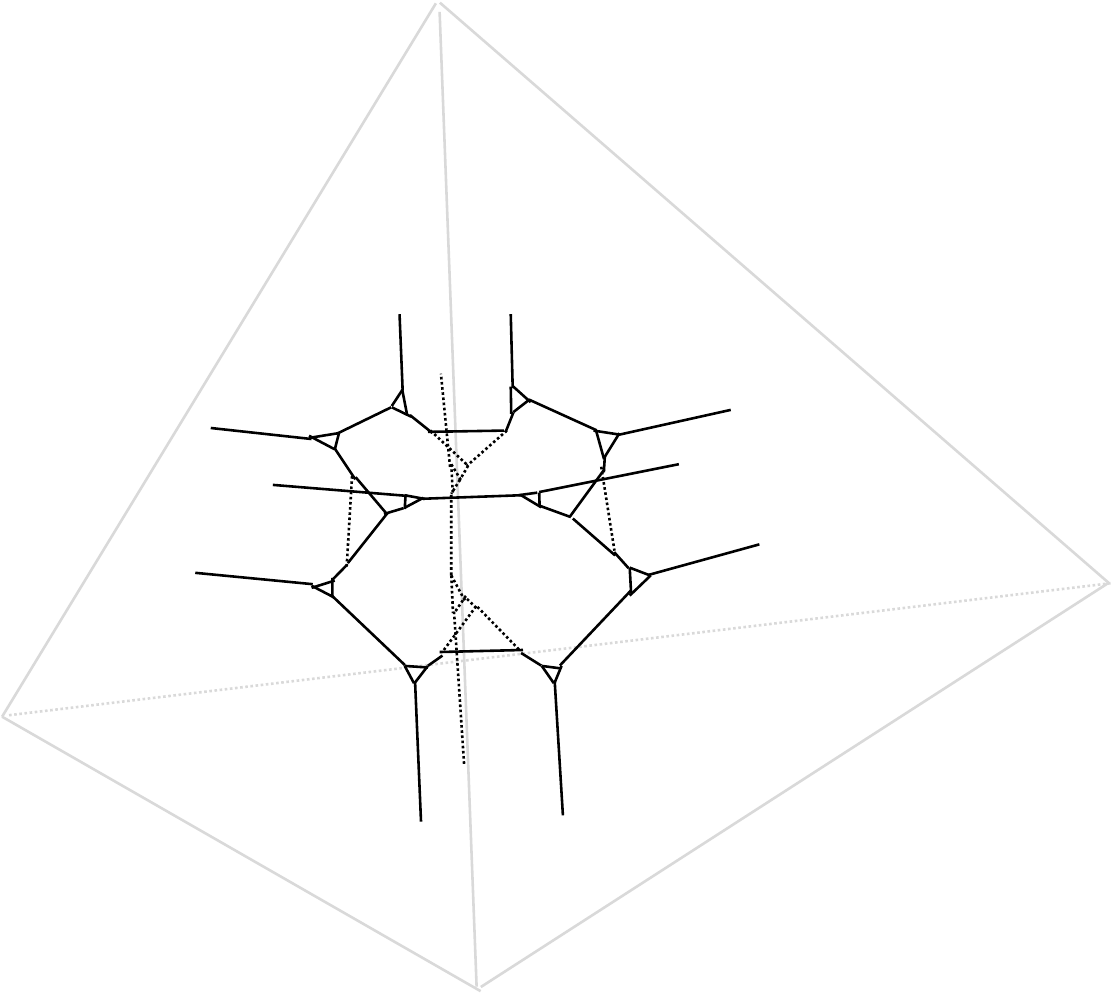}
\caption{$\Gamma$ on a 3-simplex of $\mathcal{T}$.}
\label{fig:gamma}
\end{figure}

\begin{figure}[htb]
\centering
\includegraphics{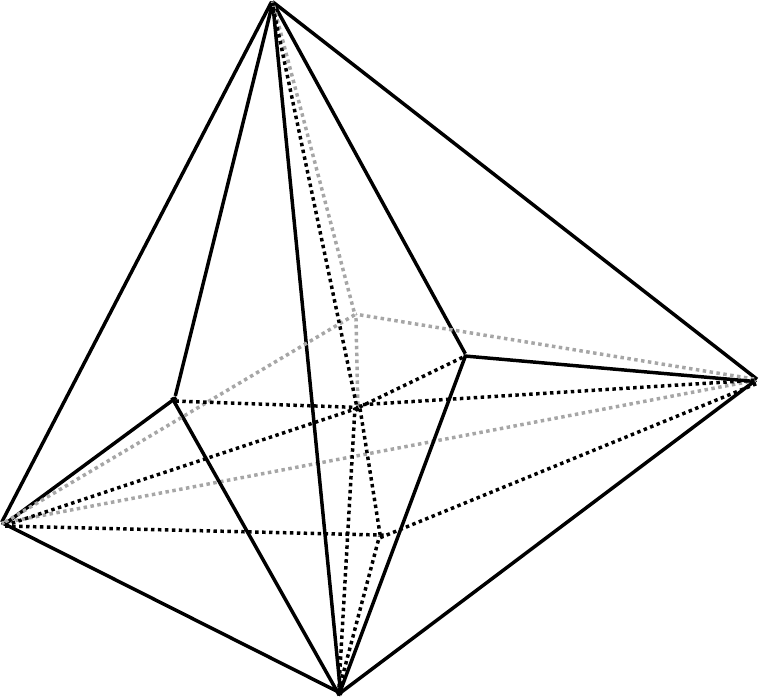}
\caption{Subdividing each 3-simplex of $\mathcal{T}$ into 12 simplices yields $\mathcal{T}'$.}
\label{fig:refined}
\end{figure}

\begin{figure}[htb]
\centering
\includegraphics{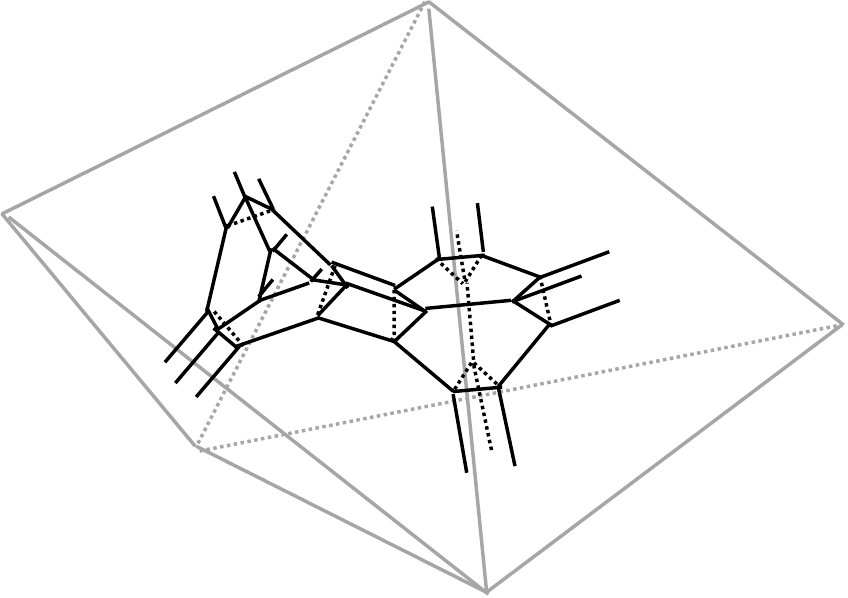}
\caption{$\Lambda$ is obtained by connecting truncated tetrahedra on 3-simplices of $\mathcal{T}$, by triangular prisms.}
\label{fig:lambda}
\end{figure}

\begin{figure}[htb]
\centering
\includegraphics{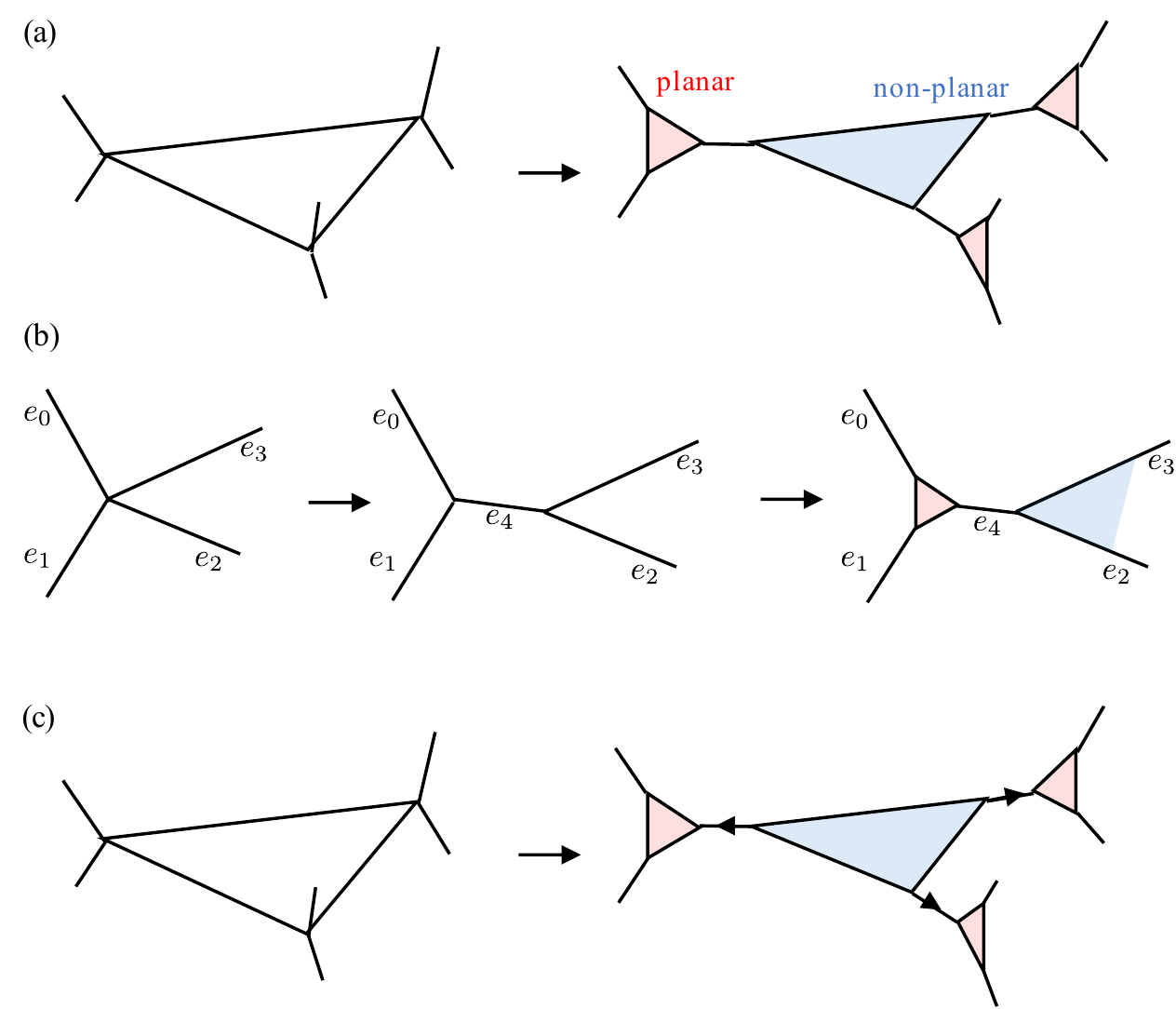}
\caption{The figure shows the process of obtaining $\Gamma$ from $\Lambda$. (a), (b): The left figure represents a triangular face of $\Lambda$.
$\Gamma$ is given by resolving each degree 4 vertex into trivalent vertices and then replacing one vertex with a triangle. A planar (resp.~non-planar) triangle is represented as a red (resp.~blue) triangle. (c): When we direct edges of $\Gamma$, the process of (a) is associated with directing newly added short edges.}
\label{fig:resolve}
\end{figure}

The trivalent graph $\Gamma$ is given by connecting patterns illustrated in Fig.~\ref{fig:gamma} on each 3-simplex of $\mathcal{T}$. For later convenience, we illustrate the following way to obtain $\Gamma$ step by step:

\begin{enumerate}
    \item First, we consider a simplical complex $\mathcal{T}'$ obtained by further subdividing each 3-simplex of $\mathcal{T}$ into 12 simplices as described in Fig.~\ref{fig:refined}. The set of 2-simplices of $\mathcal{T}'$ is denoted as $\tilde{\Lambda}$. Then, we take the dual lattice $\Lambda$ of $\tilde{\Lambda}$ described in Fig.~\ref{fig:lambda}.

    \item All vertices of $\Lambda$ have degree 4.
    For each vertex of $\Lambda$, we first resolve the vertex into a pair of trivalent vertices, and then change one of the trivalent vertices (which is not a vertex of a triangle) into three trivalent vertices, see Fig.~\ref{fig:resolve}. 
    The obtained trivalent graph is $\Gamma$ in Fig.~\ref{fig:gamma}. 
\end{enumerate}

Following the notations in Sec.~\ref{sec:tarantino}, we refer to triangular faces in $\Gamma$ as triangles. Let $t(v), t(w)$ be triangles in $\Gamma$ which contain vertices $v, w$ respectively. If an edge $\langle vw\rangle$ satisfies $t(v)=t(w)$, we refer to $\langle vw\rangle$ as a long edge. If we have $t(v)\neq t(w)$, we refer to $\langle vw\rangle$ as a short edge.
Then, the degrees of freedom in our model are given as follows:
\begin{itemize}
    \item A qubit located on each vertex of $\mathcal{T}'$, which is operated by Pauli operators $\sigma^x, \sigma^y, \sigma^z$.
    We sometimes call these qubits ``$\sigma$ qubits''. Dually, we can also think of $\sigma$ qubits as living on 3d cells of $\Gamma$.
    \item A qubit located on each 1-simplex of $\mathcal{T}'$, except for 1-simplices connecting a barycenter of a 2-simplex and a 3-simplex of $\mathcal{T}$. This qubit is operated by Pauli operators $\tau^x, \tau^y, \tau^z$, and we sometimes call these qubits ``$\tau$ qubits''. Dually, we can also think of $\tau$ qubits as living on faces of $\Gamma$ except for triangles.
    \item A pair of complex fermions located on each short edge $e$, created and annihilated by $c_{e}^{s\dagger}, c_{e}^{s}$ ($s=\uparrow, \downarrow$) respectively.
\end{itemize}

\begin{figure}[htb]
\centering
\includegraphics{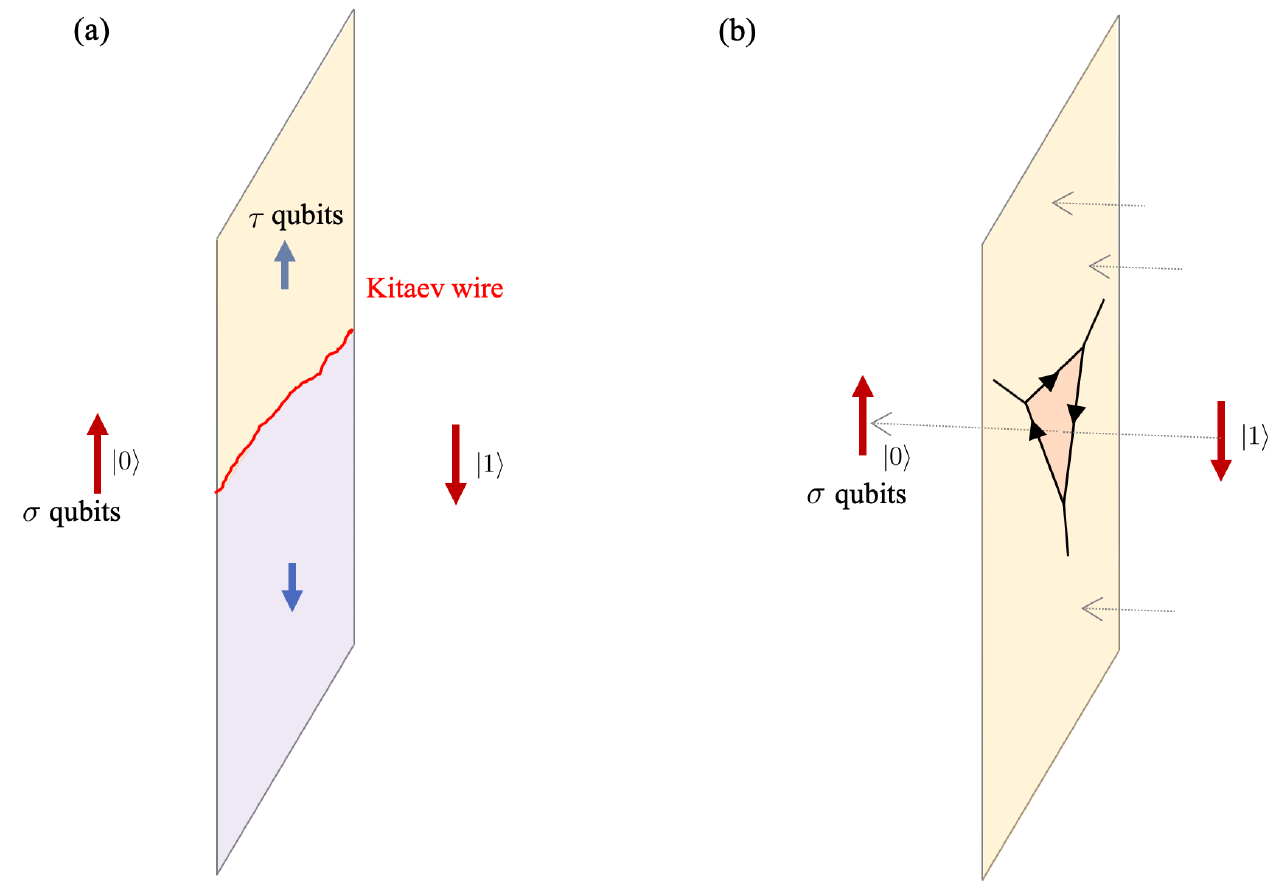}
\caption{(a): We decorate the Kitaev wire on the domain wall of $\tau$ qubits placed on the 2d domain wall of $\sigma$ qubits. (b): The section $NK$ of the normal bundle of the 2d domain wall is directed from the side of the $\ket{1}$ domain of $\sigma$ qubits to that of the $\ket{0}$ domain (gray arrows). 
Then, on each planar triangle on $\mathrm{K}$ (pink triangle), we assign directions on edges bounding the triangle clockwise around the axis parallel to the section. }
\label{fig:3ddeco}
\end{figure}

Both $\sigma$ and $\tau$ qubits are charged under time reversal as the Pauli $x$,
\begin{align}
    T: \ket{1}\mapsto\ket{0}, \ket{0}\mapsto\ket{1}.
    \label{eq:tqubit}
\end{align}
Since $T^2=(-1)^F$, fermions are also acted upon by time reversal in a nontrivial fashion. Before discussing the symmetry property of fermions, let us outline how we perform the domain wall decoration. 
Since $\sigma$ qubits are located on 3d cells of $\Gamma$, their configuration specifies a 2d domain wall on $\Gamma$, which forms a graph supported on a 2d surface, as we will define later in Sec.~\ref{subsec:wall}. Then, the configuration of $\tau$ qubits on the 2d domain wall further gives us a 1d domain wall, where we decorate the Kitaev wire; see Fig.~\ref{fig:3ddeco}. 

\subsection{Directions of edges and discrete spin structure}
\label{sec:direction}

Analogously to the Tarantino-Fidkowski type wave function in Sec.~\ref{sec:tarantino}, 
we need Kasteleyn directions on edges of $\Gamma$ restricted to the 2d domain wall of $\sigma$ qubits, in order to ensure the conservation of the fermion parity under domain wall fluctuations. Let us assume that we have obtained a 2d graph $\mathrm{K}$ on the 2d domain wall $K$, whose edges will be directed in a Kasteleyn fashion.
A caveat is that the assignment of Kasteleyn direction on $\mathrm{K}$ depends on how we choose the section of the normal bundle $NK$ of the 2d domain wall $K$; for instance, we can choose the section of $NK$ directed from the side of $\ket{1}$ domain of $\sigma$ qubits to that of $\ket{0}$ domain. Then, the Kasteleyn property is defined by the number of clockwise directed edges on a closed path of $\mathrm{K}$, around the axis parallel to the section of $NK$ (see Fig.~\ref{fig:2dgraph} (b)). 
We note that such defined Kasteleyn direction on $\mathrm{K}$ is not necessarily invariant under time reversal, since the section of $NK$ is reversed by time reversal, thereby transforming the definition of the Kasteleyn property on $\mathrm{K}$; clockwise edges now become anticlockwise.

The above observation implies that the time reversal symmetry also acts on the directions of edges. In this subsection, we will first introduce directions that are invariant under time reversal, and then discuss non-invariant directions. 
For later convenience, we classify triangles (i.e., triangular faces) in $\Gamma$ into two types: a ``non-planar'' triangle which originates from a triangular face of $\Lambda$, and a ``planar'' triangle obtained by replacing a trivalent vertex in Fig.~\ref{fig:resolve} (a). 

\subsubsection{Invariant directions on edges}
\label{sec:invarrow}
\begin{figure}[htb]
\centering
\includegraphics{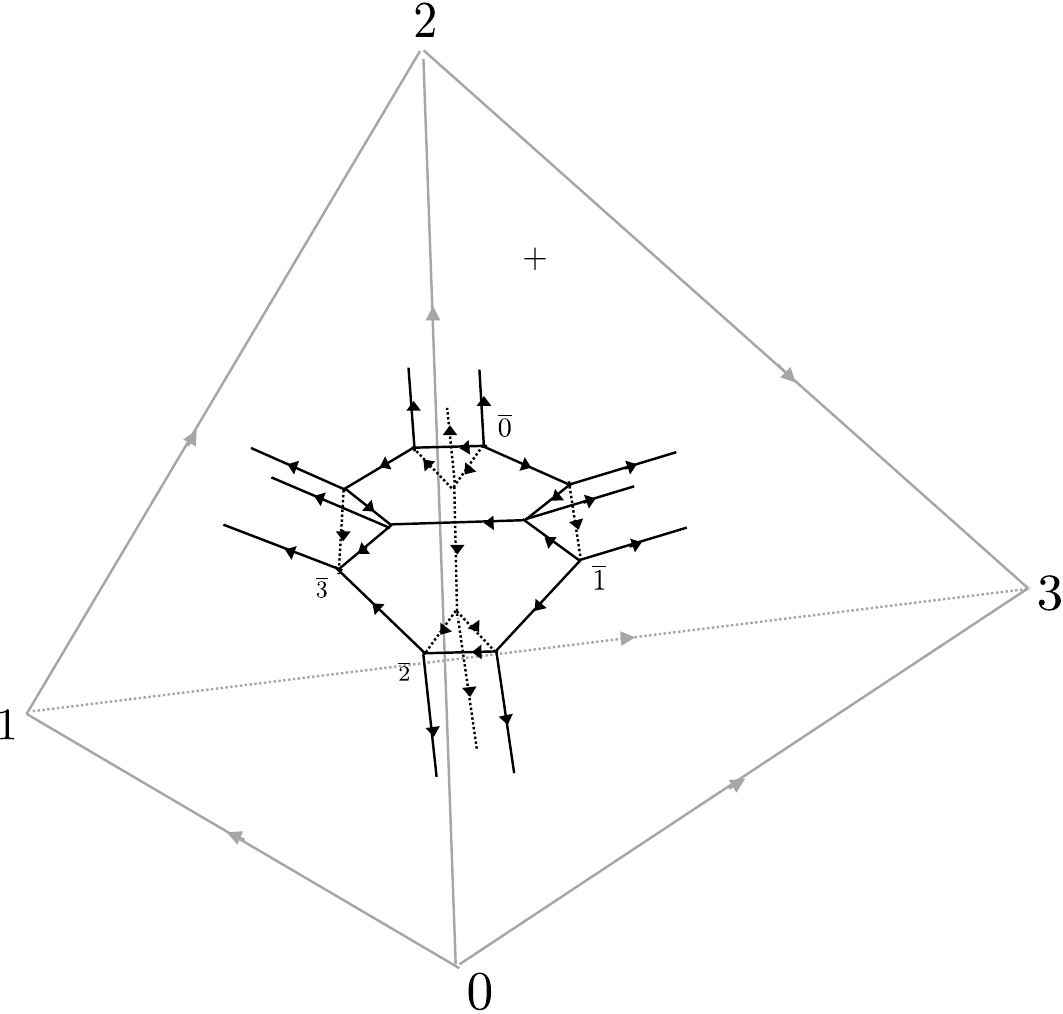}
\caption{Initial assignment of directions on edges of $\Lambda$ in a $+$ simplex.}
\label{fig:plus}
\end{figure}

\begin{figure}[htb]
\centering
\includegraphics{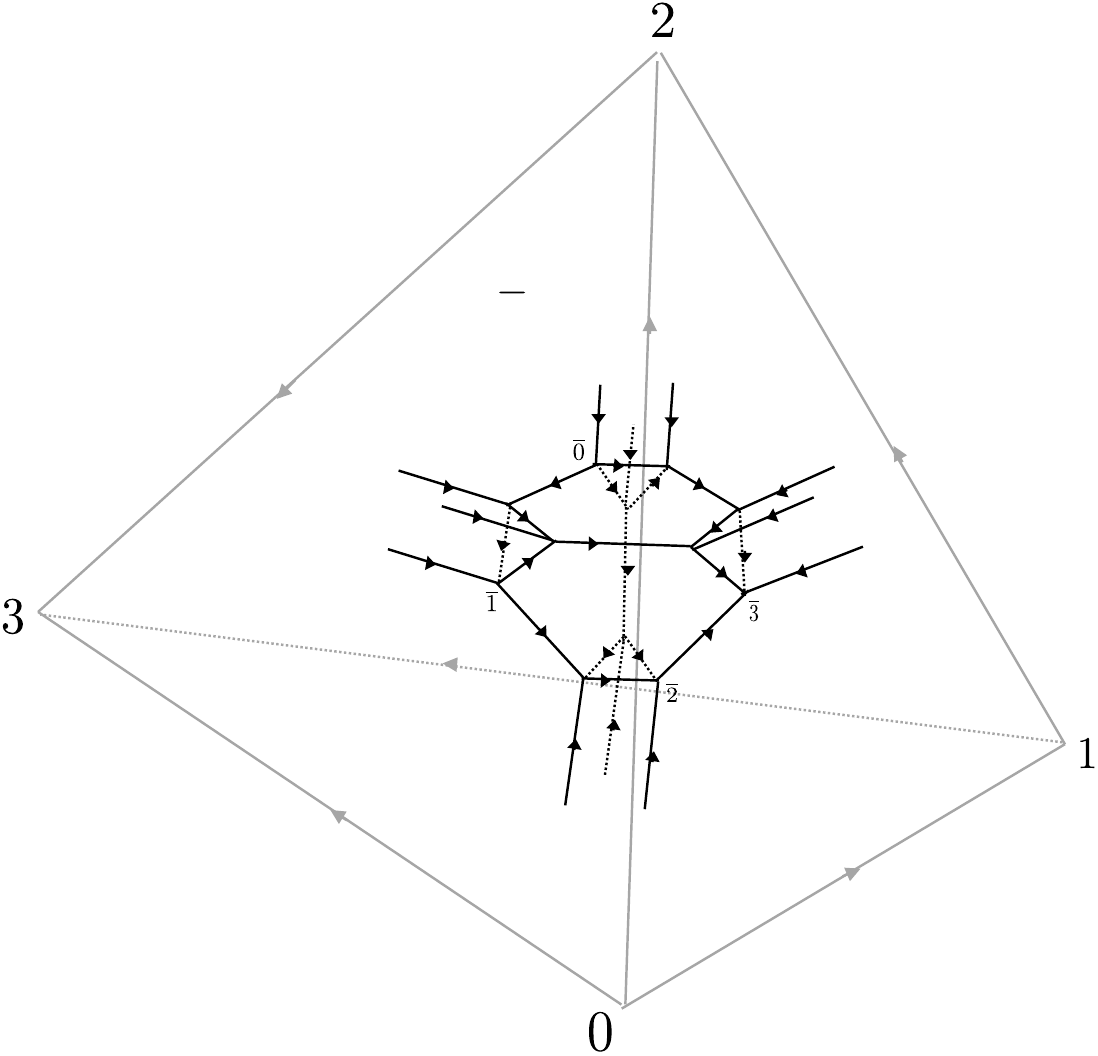}
\caption{Initial assignment of directions on edges of $\Lambda$ in a $-$ simplex.}
\label{fig:minus}
\end{figure}

Here, we introduce directions on edges of $\Gamma$ that are invariant under time reversal, determined independently of the configuration of qubits. 
These invariant directions are assigned on edges of $\Gamma$ except for edges bounding a planar triangle. Now we provide invariant directions step by step;
\begin{enumerate}
    \item We start by assigning directions on edges of the graph $\Lambda$ (Fig.~\ref{fig:lambda}), as described in Fig.~\ref{fig:plus} (resp.~Fig.~\ref{fig:minus}) for $+$ simplices (resp.~$-$ simplices) of $\mathcal{T}$.
    \item Next, we modify the directions on edges of $\Lambda$, according to the combinatorial spin structure on $M$. To define the spin structure of $M$, we first prepare the representative of the dual of the second Stiefel-Whitney class $w_2$ on the simplical complex $\mathcal{T}'$ (see Fig.~\ref{fig:refined}).
    It is represented by a 1-cycle $S\in Z_1(M,\mathbb{Z}_2)$,
    \begin{align}
    S = \sum_{e\in\mathcal{T}'} e - \sum_{\Delta_+}(\langle v_{012}v_{0123}\rangle + \langle v_{023}v_{0123}\rangle) - \sum_{\Delta_-}(\langle v_{013}v_{0123}\rangle + \langle v_{123}v_{0123}\rangle),
    \label{eq:w2formula}
    \end{align}
    where $e$ is a 1-simplex of $\mathcal{T}'$, and the first sum runs over all 1-simplices of $\mathcal{T}'$. The second sum is over 1-simplices of $\mathcal{T}'$ contained in a $+$ simplex $\Delta_+ = \langle 0123\rangle$ of $\mathcal{T}$.
    Here, the vertices of $\mathcal{T}'$ which is a barycenter of a simplex $\Delta\in\mathcal{T}$ are written as $v_{\Delta}$. Similarly, the third sum is over $-$ simplices of $\mathcal{T}$. The validity of the expression~\eqref{eq:w2formula} is proven in Appendix~\ref{app:w2}.
    The spin structure is specified by a trivialization $\partial E=S$ of $S$. Here, $E\in C_2(M,\mathbb{Z}_2)$ is a subcomplex of $\tilde{\Lambda}$. 
    
    Then, we reverse the directions of edges of $\Lambda$ that cross 2-simplices of $E$.
    
    \item Finally, we complete the assignment of directions on $\Gamma$, by generating $\Gamma$ from $\Lambda$ associated with directions to newly added short edges, as described in Fig.~\ref{fig:resolve} (c). 
\end{enumerate}

\subsubsection{Non-invariant directions on edges}
Here, we introduce directions on yet undirected edges of $\Gamma$ that are acted upon by time reversal in a nontrivial fashion. As we will see in Sec.~\ref{subsec:wall}, there will be a 2d graph $\mathrm{K}$ supported on the 2d domain wall of $\sigma$ qubits. 
We assign directions on edges bounding planar triangles, \textit{iff} the planar triangle is contained in the 2d domain wall $\mathrm{K}$ (see Fig.~\ref{fig:3ddeco} (b)). We do not assign directions if the planar triangles are away from $\mathrm{K}$.

On the 2d domain wall $K$, we choose the section of the normal bundle $NK$, such that the section of $NK$ is directed from the side of $\ket{1}$ domain of $\sigma$ qubits to that of $\ket{0}$ domain. Then, on each planar triangle on $\mathrm{K}$, we assign directions on edges bounding the triangle clockwise around the axis parallel to the section; see Fig.~\ref{fig:3ddeco} (b). 
These directions are reversed by the time reversal action, since the chosen section of $NK$ is flipped by time reversal.

\subsection{Kasteleyn direction on the 2d domain wall}
\label{subsec:wall}
Here, we define the 2d graph $\mathrm{K}$ on the 2d domain wall of $\sigma$ qubits, on which the Kasteleyn direction will be induced. 

As a technical detail, we fix each $\sigma$ qubit on the barycenter of a 2-simplex of $\mathcal{T}$ according to the majority rule: if the $\sigma$ qubit is located on the barycenter of a 2-simplex $\langle v_0v_1v_2\rangle$, it is $\ket{1}$ or $\ket{0}$ depending on whether the majority of three $\sigma$ qubits on vertices $v_0,v_1,v_2$ have $\ket{1}$ or $\ket{0}$. 
Each $\sigma$ qubit on the barycenter of a 3-simplex $\langle v_0v_1v_2v_3\rangle$ of $\mathcal{T}$ is also determined by the majority rule: it is $\ket{1}$ or $\ket{0}$ depending on whether the majority of four $\sigma$ qubits on vertices $v_0,v_1,v_2,v_3$ have $\ket{1}$ or $\ket{0}$, if the numbers of $\ket{1}$ differs from that of $\ket{0}$. If the number of $\ket{1}$ and $\ket{0}$ on $v_0,v_1,v_2,v_3$ are both 2, we leave the $\sigma$ qubit on $\langle v_0v_1v_2v_3\rangle$ undetermined.

\begin{figure}[htb]
\centering
\includegraphics{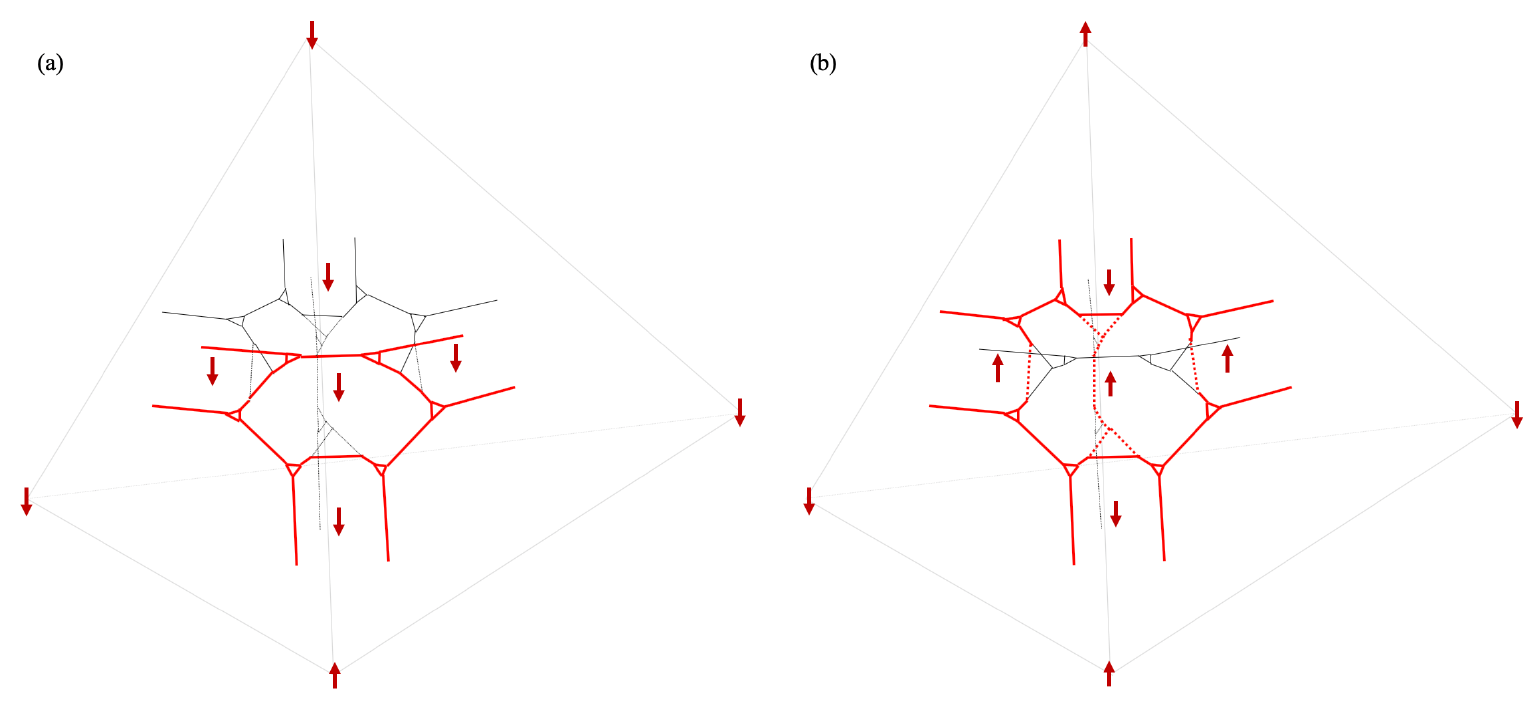}
\caption{Two possible patterns of the 2d graph $\mathrm{K}'$ in a single 3-simplex of $\mathcal{T}$, where $\sigma$ qubits are drawn as red arrows and $\mathrm{K}'$ is represented as a red graph. For (a), it is also possible to have the situation in which all $\sigma$ qubits are flipped. For (b), two non-planar triangles are contained in $\mathrm{K}'$, which are not necessarily Kasteleyn directed.}
\label{fig:2dgraph}
\end{figure}

Then, a 2d graph $\mathrm{K}'$ is defined according to the configuration of $\sigma$ qubits, as described in Fig.~\ref{fig:2dgraph}.
After some efforts, we can see that the 2d graph $\mathrm{K}'$ is ``almost Kasteleyn directed'' when seen from the side of the $\ket{1}$ domain of $\sigma$ qubits, except for non-planar triangles contained in $\mathrm{K}'$ in Fig.~\ref{fig:2dgraph} (b).

\begin{figure}[htb]
\centering
\includegraphics{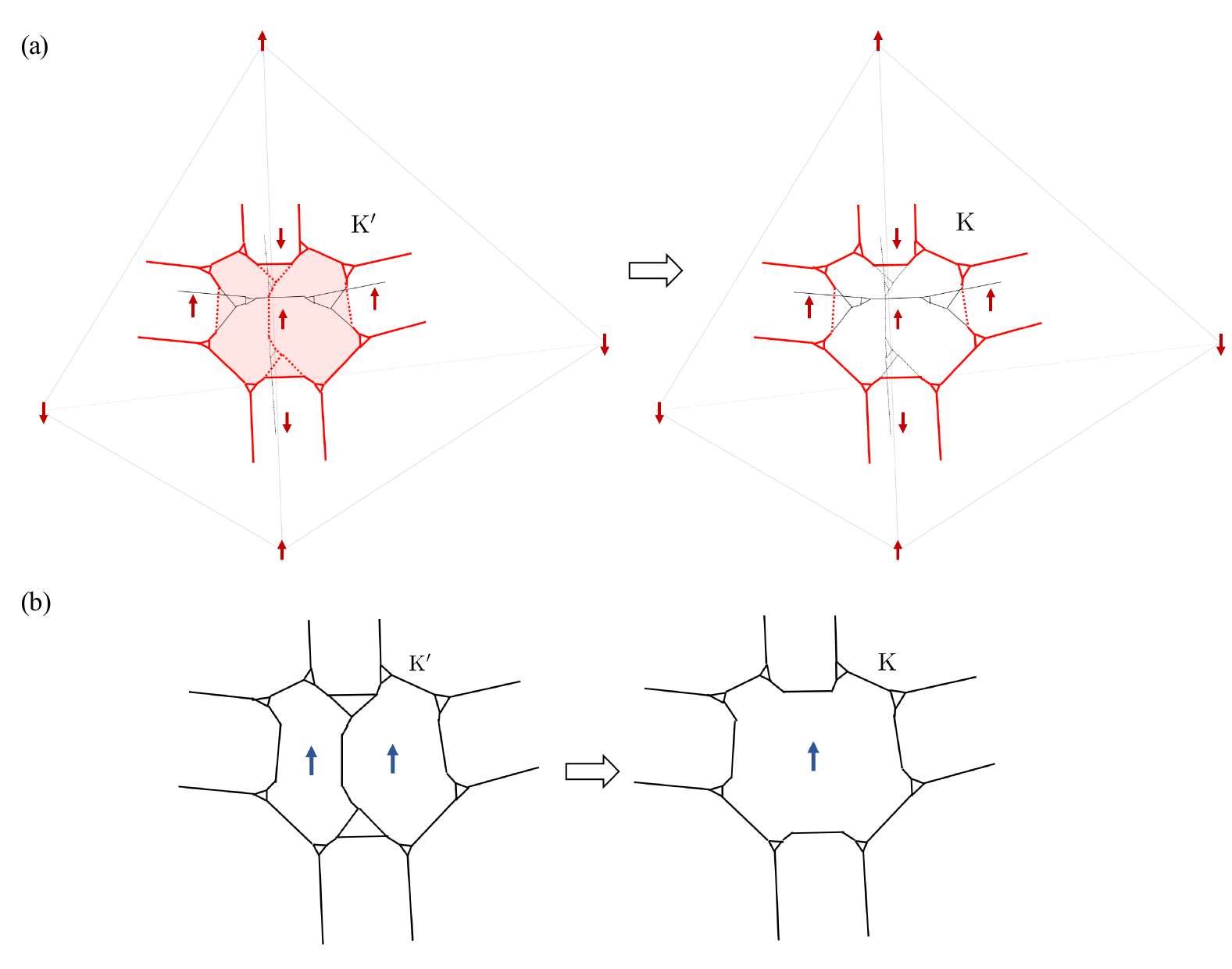}
\caption{(a): The process of obtaining $\mathrm{K}$ from $\mathrm{K}'$. This is done by gathering the four pink faces of $\mathrm{K}'$ including two non-planar triangles into a single face. (b): Two $\tau$ qubits on the identified faces of $\mathrm{K}'$ should share the same state.}
\label{fig:kpk}
\end{figure}

To prepare a graph whose edges are completely Kasteleyn directed, we gather four faces of $\mathrm{K}'$ in Fig.~\ref{fig:2dgraph} (b) into a single face, as described in Fig.~\ref{fig:kpk}. We denote the obtained graph as $\mathrm{K}$, which is completely Kasteleyn directed.

\subsection{Wave function: decorated 1d domain wall on the 2d domain wall}
\label{subsec:wfn}
Here, we precisely describe the Kitaev wire decoration on the domain wall of $\tau$ qubits on $\mathrm{K}$, which was schematically illustrated in Fig.~\ref{fig:3ddeco} (a). The decoration is based on the Kasteleyn direction on $\mathrm{K}$ introduced in the previous subsection.

In our model, we have two complex fermions on each short edge of $\Gamma$. Analogously to the (2+1)d case in Sec.~\ref{sec:tarantino}, we will represent each complex fermion on a short edge $\langle vv'\rangle$ in terms of a pair of Majorana fermions placed on vertices $v, v'$, whose assignment depends on the direction of $\langle vv'\rangle$.
Let $e = \langle \overrightarrow{vv'}\rangle$ be a short edge oriented from $v$ to $v'$. Then, each complex fermion on $e$ is represented by a pair of Majorana fermions
\begin{align}
    \begin{split}
        a_{v}^{s} &= c_{e}^{s\dagger}+c_{e}^{s}, \\
        b_{v'}^{s} &= i(c_{e}^{s\dagger}-c_{e}^{s}),
    \end{split}
    \label{eq:3dmajorana}
\end{align}
located on $v$ and $v'$, respectively. Then, we introduce the symmetry property of fermions.
Since $T^2=(-1)^F$, the fermion on $\langle \overrightarrow{vv'}\rangle$ is a Kramers doublet under $T$,
\begin{align}
T:
\begin{cases}
a_v^{\uparrow}\to a_v^{\downarrow}\\
a_v^{\downarrow}\to-a_v^{\uparrow},
\end{cases}
\begin{cases}
b_{v'}^{\uparrow}\to-b_{v'}^{\downarrow}\\
b_{v'}^{\downarrow}\to b_{v'}^{\uparrow}.
\end{cases}
\label{eq:tfermion}
\end{align}

The wave function of our model is given by pairing up Majorana fermions on vertices, according to a dimer configuration on $\Gamma$.  
Similar to the (2+1)d case in Sec.~\ref{sec:tarantino}, away from the 2d domain wall $\mathrm{K}$, we pair up Majorana fermions that share a short edge $\langle \overrightarrow{vv'}\rangle$, by a pairing term $ia_v^{\uparrow}b_{v'}^{\uparrow}+ia_v^{\downarrow}b_{v'}^{\downarrow}$.
Furthermore, the $\tau$ qubits on the face of $\Gamma$ are fixed away from $\mathrm{K}$, depending on the domain of $\sigma$ qubits: the $\tau$ qubits are $\ket{1}$ (resp.~$\ket{0}$) if contained in the domain of $\ket{1}$ (resp.~$\ket{0}$).

Next, we consider the domain wall of $\tau$ qubits on $\mathrm{K}$. 
As a technical detail, we recall that the 2d graph $\mathrm{K}$ was obtained by gathering four faces in $\mathrm{K}'$ into a single face, which was described in Fig.~\ref{fig:kpk}. 
Since we have one $\tau$ qubit on each face of $\mathrm{K}'$ except for triangles, the newly obtained single face of $\mathrm{K}$ in Fig.~\ref{fig:kpk} (a) contains two $\tau$ qubits. 
To consider the Kitaev wire decoration on $\mathrm{K}$ instead of $\mathrm{K}'$, we have to make sure that two $\tau$ qubits share the same state, i.e., $\ket{00}$ or $\ket{11}$, as described in Fig.~\ref{fig:kpk} (b).~\footnote{
This is done by introducing a term e.g., $-\tau^z_f\tau^z_{f'}$, where $f, f'$ denote two faces of $\mathrm{K}'$ gathered in Fig.~\ref{fig:kpk}.} 

Then, away from the domain wall of $\tau$ qubits on $\mathrm{K}$, we also pair up Majorana fermions that share a short edge $\langle \overrightarrow{vv'}\rangle$, by $ia_v^{\uparrow}b_{v'}^{\uparrow}+ia_v^{\downarrow}b_{v'}^{\downarrow}$. These pairings away from the Kitaev wire decoration are invariant under $T$, which is consistent with the fact that the directions of short edges are unchanged by time reversal, according to Sec.~\ref{sec:direction}.

\subsubsection{Kitaev wire on the 1d domain wall}
\begin{figure}[htb]
\centering
\includegraphics{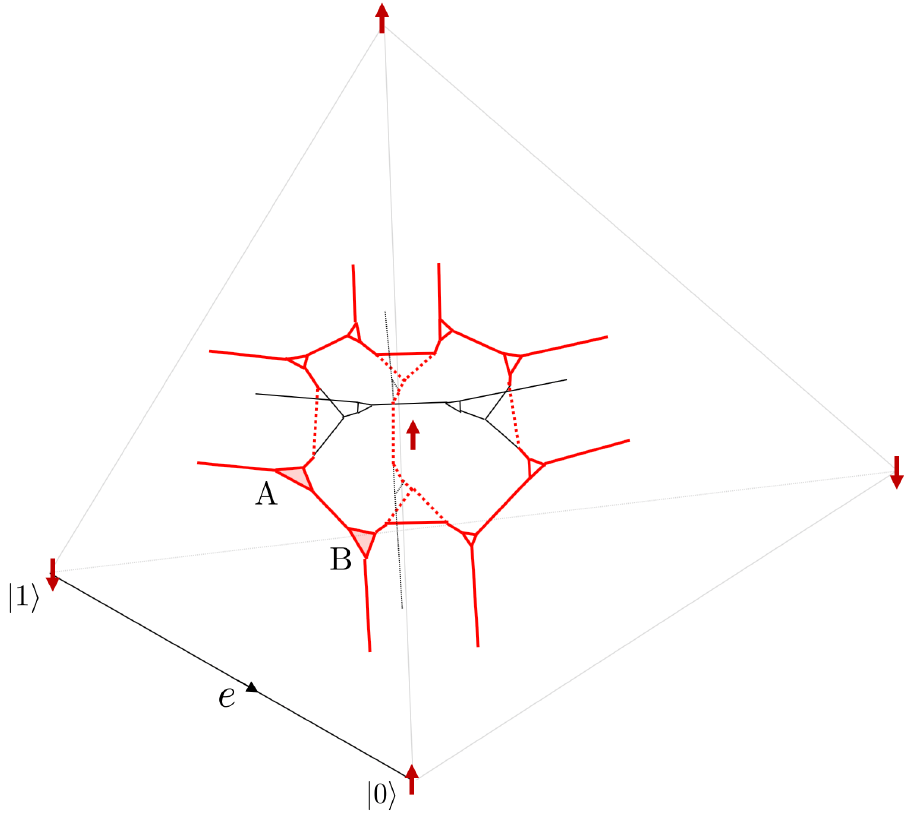}
\caption{
Labeling all planar triangles on $\mathrm{K}$ with either ``A'' or ``B''. For a 1-simplex $e$ of $\mathcal{T}$ crossing the 2d domain wall, we find a pair of planar triangles contained in a single 3-simplex of $\mathcal{T}$, which is located in the nearest position of $e$. 
Then, we label the pair of planar triangles by ``A'' and ``B'', such that the ``A'' triangle is located in the clockwise direction of the ``B'' triangle, when seen from the side of $\ket{1}$ domain of $\sigma$ qubits.}
\label{fig:AB}
\end{figure}

Now we explain the way to put the Kitaev wire on the 1d domain wall of $\tau$ qubits on $\mathrm{K}$. Analogously to the (2+1)d case in Sec.~\ref{sec:tarantino}, this is done by pairing Majorana fermions along the long edges on the 1d domain wall. 
To do this, it is convenient to label the planar triangle on $\mathrm{K}$ in a ``bipartite'' fashion. For a 1-simplex $e$ of $\mathcal{T}$ crossing the 2d domain wall, we find a pair of planar triangles contained in a single 3-simplex of $\mathcal{T}$, which is located in the nearest position of $e$, as described in Fig.~\ref{fig:AB}.
Then, we label the pair of planar triangles by ``A'' and ``B'', such that the ``A'' triangle is located in the clockwise direction of the ``B'' triangle, when seen from the side of $\ket{1}$ domain of $\sigma$ qubits.

Then, on the 1d domain wall $\tau$ qubits on $\mathrm{K}$, we pick out Majorana modes $\gamma_v^{s_v}$ ($\gamma$ is $a$ or $b$) from each vertex $v$, and we pair them along the long edges bounding a planar triangle $\langle \overrightarrow{vw}\rangle$ as $i\gamma_v^{s_v}\gamma_{w}^{s_w}$, so that they form the Kitaev wire. $s_v$ and $s_w$ are determined on each vertex according to the following rule,
\begin{itemize}
\item If the planar triangle is labeled by ``A'', $s=\ \uparrow$ when $\gamma$ is $a$, and $s=\ \downarrow$ when $\gamma$ is $b$.
\item If the planar triangle is labeled by ``B'', $s=\ \downarrow$ when $\gamma$ is $a$, and $s=\ \uparrow$ when $\gamma$ is $b$.
\end{itemize}

After pairing up these Majorana fermions, we are left with one unpaired Majorana on each vertex of planar triangles, and two on each vertex of non-planar triangles, on the 1d domain wall. 
Then, we pair up yet unpaired Majorana fermions on short edges $\langle \overrightarrow{vw}\rangle$, as $i\gamma_v^{\overline{s}_v}\gamma_{w}^{\overline{s}_w}$. 
Here, we will choose the pairing such that $\overline{s}_v=\overline{s}_w$.

Finally, we have one unpaired Majorana fermion on each vertex of non-planar triangles. We pair them up along long edges bounding a non-planar triangle $\langle \overrightarrow{vw}\rangle$ as $i\gamma_v^{s_v}\gamma_{w}^{s_w}$. Here, we can see that $s_v$ is flipped from $s_w$, $s_v=-s_w$ (here, $-s$ denotes the opposite spin to $s$).

\subsubsection{Wave function}
The wave function of our model is given by the equal superposition of all possible configurations of $\sigma$ qubits and $\tau$ qubits, associated with the Kitaev wire decoration discussed above. 

Let us demonstrate the time reversal invariance of this wave function. To see this, we first note that the pairing of Majorana fermions away from the Kitaev wire decoration $ia_v^{\uparrow}b_{v'}^{\uparrow}+ia_v^{\downarrow}b_{v'}^{\downarrow}$ is invariant under time reversal. On the Kitaev wire decoration, according to the pairing rule, the spins of paired Majorana fermions $s_v, s_w$ flip their signs under time reversal, which is consistent with the transformation law of fermions. This is because the labels of planar triangles (``A'' or ``B'') are changed under time reversal, thereby the spins of paired Majorana fermions are also flipped.
We can also check that the pairings of Majorana fermions are consistent with the induced Kasteleyn direction on $\mathrm{K}$ under time reversal. 
On one hand, on short edges and long edges bounding a non-planar triangle on $\mathrm{K}$, the sign of the pairing $i\gamma_v^{s_v}\gamma_{w}^{s_w}$ is invariant under time reversal, which is consistent with the invariance of the direction on $\langle vw\rangle$. 
On the other hand, on long edges of planar triangles of $\mathrm{K}$, the pairing $i\gamma_v^{s_v}\gamma_{w}^{s_w}$ flips its sign under time reversal. 
It is also consistent with the Kasteleyn direction on $\mathrm{K}$, which flips the directions on long edges of planar triangles.

Thanks to the Kasteleyn directions induced on the 2d domain walls, the obtained wave function also preserves the $\mathbb{Z}_2^F$ symmetry, analogously to the (2+1)d case in Sec.~\ref{sec:tarantino}.

\subsection{Hamiltonian}
\label{sec:3dham}
Now we provide the Hamiltonian for our (3+1)d model. 
The Hamiltonian consists of four terms,
\begin{align}
    H= H_{\mathrm{qubit}}+H_{\mathrm{decorate}}+H_{\tau\mathrm{fluct}}+H_{\sigma\mathrm{fluct}}.
    \label{eq:ham}
\end{align}

\subsubsection{$H_{\mathrm{qubit}}$}

The first term $H_{\mathrm{qubit}}$ is defined to stabilize the desired state of $\sigma$ and $\tau$ qubits, which does not involve any fermionic operator. Summarizing the properties of the ground state illustrated in the previous subsections, state with the following properties should be realized;

\begin{itemize}
    \item Each $\sigma$ qubit on the barycenter of a 2-simplex $\langle v_0v_1v_2\rangle$ of $\mathcal{T}$ is fixed according to the majority rule.
    
    \item Each $\sigma$ qubit on the barycenter of a 3-simplex $\langle v_0v_1v_2v_3\rangle$ of $\mathcal{T}$ is also determined by the majority rule. 
    
    \item Each $\tau$ qubit on the face of $\Gamma$ is fixed if it is away from $\mathrm{K}$, depending on the domain of $\sigma$ qubits: $\ket{1}$ (resp.~$\ket{0}$) if contained in the domain of $\ket{1}$ (resp.~$\ket{0}$).
    
    \item The 2d graph $\mathrm{K}$ was obtained by gathering four faces in $\mathrm{K}'$ into a single face, which was described in Fig.~\ref{fig:kpk}. 
    Since we have one $\tau$ qubit on each face of $\mathrm{K}'$ except for triangles, the newly obtained single face of $\mathrm{K}$ in Fig.~\ref{fig:kpk} contains two $\tau$ qubits.
    These two $\tau$ qubits on faces $f, f'$ of $\mathrm{K}'$ share the same state, i.e., $\ket{00}$ or $\ket{11}$. 
\end{itemize}
All four of these properties are realized by preparing a local Hamiltonian, which is represented as a polynomial of $\sigma^z, \tau^z$ operators. The explicit form of $H_{\mathrm{qubit}}$ is given in Appendix~\ref{app:hqubit}.

\subsubsection{$H_{\mathrm{decorate}}$}

Next, let us introduce the term $H_{\mathrm{decorate}}$ in~\eqref{eq:ham}, which realizes the desired pairing of Majorana fermions as described in Sec.~\ref{subsec:wfn}.
As we have seen in Sec.~\ref{subsec:wfn}, the pairing rule of Majorana fermions (i.e., the choice of the spin $s_v, s_w$ for the pairing $i\gamma_v^{s_v}\gamma_{w}^{s_w}$) on an edge $\langle vw\rangle$ is completely determined by the value of $\sigma^z, \tau^z$ of qubits in the vicinity of $\langle vw\rangle$. 
Let $R_{\langle vw\rangle}$ be a set of $\sigma$ and $\tau$ qubits whose configuration determines the pairing rule of Majorana fermions at $\langle vw\rangle$.
Clearly,  $R_{\langle vw\rangle}$ can be taken locally from $\langle vw\rangle$.
Then, $H_{\mathrm{decorate}}$ can be expressed in the form of
\begin{align}
\begin{split}
    H_{\mathrm{decorate}} &=\sum_{\substack{\mathrm{short\ edge}\\ \langle\overrightarrow{vw}\rangle}}\sum_{\{s^z, t^z\}} (ia_v^{\uparrow}b_{w}^{\uparrow}+ia_v^{\downarrow}b_{w}^{\downarrow})(1-D_{\langle vw\rangle})P_{\mathrm{qubit}}^{\{s^z,t^z\}}+\sum_{\langle \overrightarrow{vw}\rangle} \sum_{\{s^z, t^z\}} (i\gamma_v^{s_v}\gamma_{w}^{s_w})D_{\langle vw\rangle}P_{\mathrm{qubit}}^{\{s^z,t^z\}}.
    \end{split}
    \label{eq:hdeco}
\end{align}
Here, the first term realizes the pairing $ia_v^{\uparrow}b_{w}^{\uparrow}+ia_v^{\downarrow}b_{w}^{\downarrow}$ away from the Kitaev wire, and the second term is for the pairing on the Kitaev wire.
$D_{\langle vw\rangle}$ is an operator of qubits which returns 1 when we decorate the Kitaev wire on $\langle vw\rangle$, otherwise returns 0. 
The sum over ``$\{s^z, t^z\}$'' runs over $2^{|R_{\langle vw\rangle}|}$ patterns of $\sigma^z, \tau^z$ eigenvalues $\{s^z,t^z\}$ of qubits in $R_{\langle vw\rangle}$. Then, $P_{\mathrm{qubit}}^{\{s^z,t^z\}}$ is defined as an operator of qubits that returns 1 if the set of eigenvalues $\{s^z, t^z\}$ is permitted by $H_{\mathrm{qubit}}$, otherwise 0. Specifically, $P_{\mathrm{qubit}}^{\{s^z,t^z\}}$ has the form of
\begin{align}
    P_{\mathrm{qubit}}^{\{s^z,t^z\}}=P_{\mathrm{qubit}}^{R_{\langle vw\rangle}}\prod_{\sigma\  \mathrm{qubits}\in  R_{\langle vw\rangle} }\frac{1+s^z_c\sigma^z_c}{2}\prod_{\tau\ \mathrm{qubits} \in R_{\langle vw\rangle}}\frac{1+ t^z_f\tau^z_f}{2},
\end{align}
where $P_{\mathrm{qubit}}^{R_{\langle vw\rangle}}$ is a projector of qubits in $R_{\langle vw\rangle}$ which projects onto the ground state of $H_{\mathrm{qubit}}$ supported on $R_{\langle vw\rangle}$.

\subsubsection{$H_{\tau\mathrm{fluct}}$}
$H_{\tau\mathrm{fluct}}$ will be defined to tunnel between the different $\tau$ qubit configurations, for a fixed configuration of $\sigma$ qubits. $H_{\tau\mathrm{fluct}}$ acts only on $\tau$ qubits on the 2d domain wall of $\sigma$ qubits, associated with the tunneling of the Kitaev wire restricted to the 2d domain wall. 
$H_{\tau\mathrm{fluct}}$ is defined in parallel with the case of (2+1)d in Sec.~\ref{sec:tarantino}, in the form of
\begin{align}
    H_{\tau\mathrm{fluct}}=\sum_f\tau_f^x X_f,
    \label{eq:htfluct}
\end{align}
where $X_f$ rearranges the dimer configuration of Majorana pairings, associated with the bit flip $\tau_f^x$ at the face $f$,
\begin{align}
    X_f=\sum_{\{s^z,t^z\};  R_f}X_f^{\{s^z,t^z\}}\Pi_f^{\{s^z,t^z\}}D_fP_f^{\{s^z,t^z\}}.
    \label{eq:xf3d}
\end{align}
Here, $R_f$ denotes a set of qubits whose eigenvalues of $\sigma^z, \tau^z$ determine the pairing rule of Majorana fermions on vertices of $f$. 
$R_f$ contains $\tau$ qubits on $f$ and faces adjacent to $f$ in $\mathrm{K}$, and $\sigma$ qubits sufficient to define $\mathrm{K}$ in the vicinity of $f$.
Clearly, we can set $R_f$ locally from $f$.
Then, the sum in~\eqref{eq:xf3d} is over $2^{|R_f|}$ patterns of $\sigma^z, \tau^z$ eigenvalues $\{s^z,t^z\}$ in $R_f$.  $P_{f}^{\{s^z,t^z\}}$ is a projector for qubits in $R_f$ which stabilizes a given set of eigenvalues $\{s^z, t^z\}$ of $\{\sigma^z,\tau^z\}$ in the summand. 
$D_f$ is an operator of $\sigma$ qubits which picks up the states with a 2d domain wall on the face $f$. Specifically,

\begin{align}
    P_{f}^{\{s^z,t^z\}}=P_{\mathrm{qubit}}^{R_f}\prod_{\sigma\ \mathrm{qubits}\in R_f}
    \frac{1+s^z_{c}\sigma^z_{c}}{2}\prod_{\tau\ \mathrm{qubits}\in R_f}
    \frac{1+t^z_{f'}\tau^z_{f'}}{2},
\end{align}
where $P_{\mathrm{qubit}}^{R_f}$ is a projector of qubits in $R_f$ which projects onto the ground state of $H_{\mathrm{qubit}}$ on $R_f$.
\begin{align}
    D_f=\frac{1-\sigma_c^z\sigma_{c'}^z}{2},
\end{align}
where $c, c'$ denote the 3d cells of $\Gamma$ sandwiching $f$.

\begin{figure}[htb]
\centering
\includegraphics{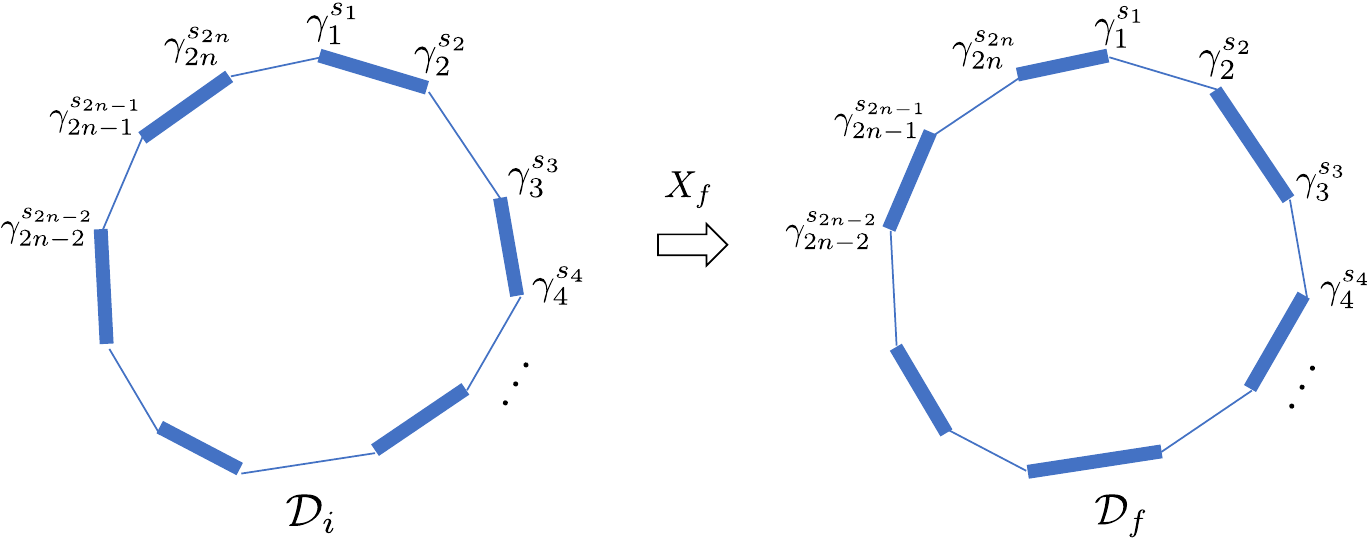}
\caption{Two different configurations $\mathcal{D}_i$, $\mathcal{D}_f$ of Majorana pairings are related by sliding dimers along a closed path on $\Gamma$.}
\label{fig:dimer}
\end{figure}

By the action of the bit flip $\tau^z_f$, the Majorana pairings are rearranged from the initial dimer configuration $\mathcal{D}_i$ to the final one $\mathcal{D}_f$. These two dimer coverings are related by sliding a sequence of dimers along a closed path $C$ of $\Gamma$. 
Suppose edges $\langle v_1v_2\rangle, \langle v_3v_4\rangle, \dots,\langle v_{2n-1}v_{2n}\rangle$ form dimers in $\mathcal{D}_i$, which are rearranged to $\langle v_2v_3\rangle, \langle v_4v_5\rangle, \dots\langle v_{2n}v_{1}\rangle$ in $\mathcal{D}_f$, as described in Fig.~\ref{fig:dimer}. 

$\Pi_f^{\{s^z,t^z\}}$ is a projector for the fermionic Hilbert space which stabilizes the Majorana pairings according to the dimer configuration $\mathcal{D}_i$.
$X_f^{\{s^z,t^z\}}$ has the effect of moving Majorana pairings along $C$. Specifically,
\begin{align}
\begin{split}
    \Pi_{f}^{\{s^z,t^z\}}=& \left(\frac{1+is_{1,2}\gamma_1^{s_1}\gamma_2^{s_2}}{2}\right)\left(\frac{1+is_{3,4}\gamma_3^{s_3}\gamma_4^{s_4}}{2}\right)\dots \left(\frac{1+is_{2n-1,2n}\gamma_{2n-1}^{s_{2n-1}}\gamma_{2n}^{s_{2n}}}{2}\right),
    \end{split}
    \label{eq:pif}
\end{align}

\begin{align}
    X_f^{\{s^z,t^z\}}=2^{-\frac{n+1}{2}}(1+is_{2,3}\gamma_2^{s_2}\gamma_3^{s_3})(1+is_{4,5}\gamma_4^{s_4}\gamma_5^{s_5})\dots(1+is_{2n,1}\gamma_{2n}^{s_{2n}}\gamma_1^{s_1}),
    \label{eq:xf}
\end{align}
where spins $s_i$ are determined by the pairing rule of Majorana fermions introduced in Sec.~\ref{subsec:wfn}. $s_{i,j}=1$ if the direction for $\langle v_iv_j\rangle$ is $\langle\overrightarrow{v_iv_j}\rangle$, and $s_{i,j}=-1$ for the opposite direction.

\subsubsection{$H_{\sigma\mathrm{fluct}}$}
\begin{figure}[htb]
\centering
\includegraphics{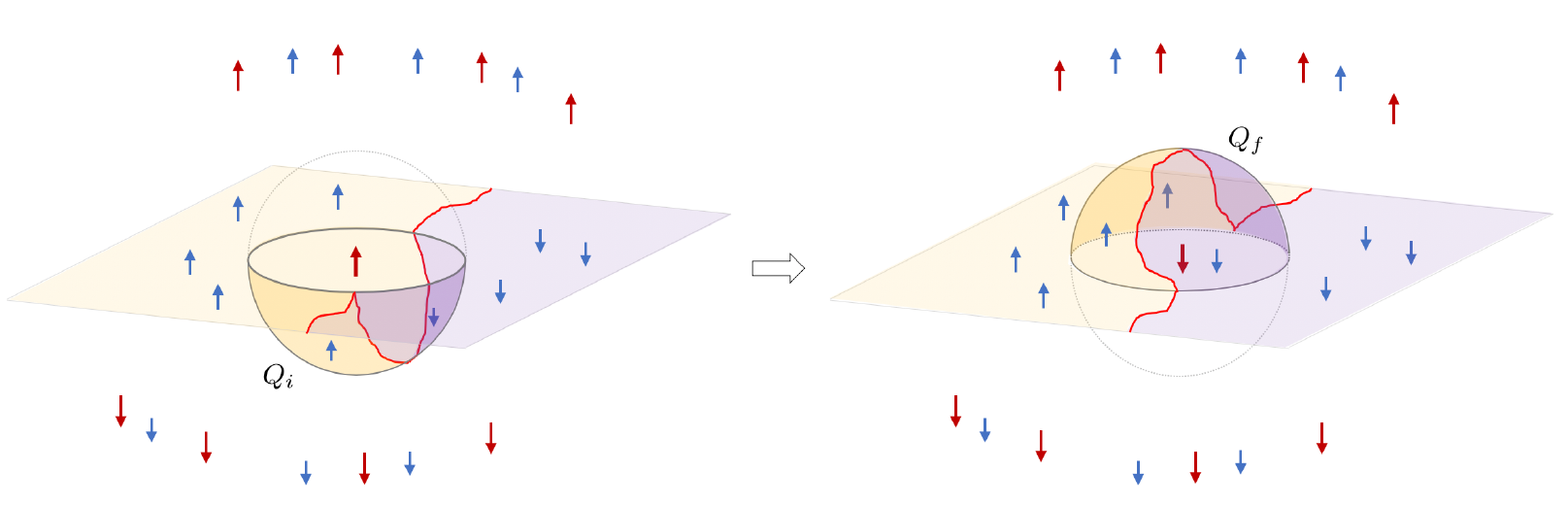}
\caption{The 2d domain wall is moved by operating the bit flip $\sigma^x$ on a single $\sigma$ qubit. A 3d cell supporting the flipped $\sigma$ qubit is represented as a sphere. In the initial (resp.~final) state, $\tau$ qubits in $Q_i$ (resp.~$Q_f$) are freely fluctuating since they are lying on the domain wall. }
\label{fig:sigmafluct}
\end{figure}

$H_{\sigma\mathrm{fluct}}$ involves the bit flip of a $\sigma$ qubit. Suppose initially we have a specific configuration of $\sigma$ and $\tau$ qubits, with a graph $\mathrm{K}_{i}$ defined on the 2d domain wall (see Fig.~\ref{fig:2dgraph}). 
We operate the bit flip $\sigma^x_c$ on a 3d cell $c$ for the initial state, which moves the 2d domain wall, providing the final graph $\mathrm{K}_{f}$ on the resulting 2d domain wall. 
We denote $Q_i$ (resp.~$Q_f$) as the set of $\tau$ qubits which are contained in faces of $\mathrm{K}_{i}$ (resp.~$\mathrm{K}_{f}$), but not contained in faces of $\mathrm{K}_{f}$ (resp.~$\mathrm{K}_{i}$). The profile of $\mathrm{K}_{i}, \mathrm{K}_{f}$ are schematically described in Fig.~\ref{fig:sigmafluct}.

Then, $H_{\sigma\mathrm{fluct}}$ is defined in the form of
\begin{align}
    H_{\sigma\mathrm{fluct}}=\sum_{c}\sigma_c^xY_c,
\end{align}
where the sum is over the 3d cells of $\Gamma$. 
On one hand, the qubits in $Q_{i}$ fluctuate freely in the initial state, since these qubits are located on the 2d domain wall (see Fig.~\ref{fig:sigmafluct}). After moving the 2d domain wall by the $\sigma_c^x$ action, the qubits in $Q_{i}$ are frozen in the state required by $H_{\mathrm{qubit}}$. 
On the other hand, the qubits in $Q_{f}$ are frozen in the initial state required by $H_{\mathrm{qubit}}$, which can freely fluctuate after moving the 2d domain wall by the action of $\sigma_c^x$ (see Fig.~\ref{fig:sigmafluct}). 
These rearrangements of $\tau$ qubits are realized by $Y_c$. Specifically, 
\begin{align}
    Y_c=2^{-\frac{|Q_f|}{2}}P_{\mathrm{qubit}}^{R_c}\cdot\sum_{S\subset Q_i\cup Q_f}\left(\prod_{\tau\ \mathrm{qubits}\in S}\tau_f^x\cdot Y_c^S\right)\cdot P_{\mathrm{qubit}}^{R_c},
\end{align}
where $R_c$ is the set of $\sigma, \tau$ qubits whose configuration determines the pairing rule of Majorana fermions on vertices of the 3d cell $c$. $P_{\mathrm{qubit}}^{R_c}$ is a projector of qubits in $R_c$ which projects onto the ground state of $H_{\mathrm{qubit}}$ supported on $R_c$. 
The sum runs over $2^{|Q_i\cup Q_f|}$ patterns of the subset $S$ of $Q_i\cup Q_f$. $Y_c^S$ rearranges the Majorana pairings according to the bit flip of $\sigma, \tau$ qubits,
\begin{align}
    Y_c^S=\sum_{\{s^z,t^z\};R_c}Y_c^{S,\{s^z,t^z\}}\Pi_c^{S,\{s^z,t^z\}}P_c^{S,\{s^z,t^z\}}.
\end{align}
The sum is over $2^{|R_c|}$ patterns of $\sigma^z, \tau^z$ eigenvalues $\{s^z,t^z\}$ in $R_c$.  $P_{c}^{S,\{s^z,t^z\}}$ is a projector for qubits in $R_c$ which stabilizes a given set of eigenvalues $\{s^z, t^z\}$ of $\{\sigma^z,\tau^z\}$ in the summand.
\begin{align}
\begin{split}
    P_c^{S,\{s^z,t^z\}}=&\left(\prod_{\tau\ \mathrm{qubits}\in S}\tau_f^x\right)P_{\mathrm{qubit}}^{R_c}\left(\prod_{\tau\ \mathrm{qubits}\in S}\tau_f^x\right) P_{\mathrm{qubit}}^{R_c} \\
    &\cdot\prod_{\sigma\ \mathrm{qubits}\in R_c}
    \frac{1+s^z_{c'}\sigma^z_{c'}}{2}\prod_{\tau\ \mathrm{qubits}\in R_c}
    \frac{1+t^z_{f}\tau^z_{f}}{2}.
    \end{split}
\end{align}
By the action of the bit flips (for a $\sigma$ qubit on $c$ and $\tau$ qubits in $S$), the Majorana pairings are rearranged from the initial dimer configuration $\mathcal{D}_i$ to the final one $\mathcal{D}_f$. These two dimer coverings are related by sliding a sequence of dimers along a closed path $C$ of $\Gamma$. 
Suppose edges $\langle v_1v_2\rangle, \langle v_3v_4\rangle, \dots,\langle v_{2n-1}v_{2n}\rangle$ form dimers in $\mathcal{D}_i$, which are rearranged to $\langle v_2v_3\rangle, \langle v_4v_5\rangle, \dots\langle v_{2n}v_{1}\rangle$ in $\mathcal{D}_f$, as described in Fig.~\ref{fig:dimer}. 

$\Pi_c^{S,\{s^z,t^z\}}$ is a projector for the fermionic Hilbert space which stabilizes the Majorana pairings according to the dimer configuration $\mathcal{D}_i$.
$Y_c^{S,\{s^z,t^z\}}$ has the effect of moving Majorana pairings along $C$. Specifically,
\begin{align}
\begin{split}
    \Pi_{c}^{S,\{s^z,t^z\}}=& \left(\frac{1+is_{1,2}\gamma_1^{s_1}\gamma_2^{s_2}}{2}\right)\left(\frac{1+is_{3,4}\gamma_3^{s_3}\gamma_4^{s_4}}{2}\right)\dots \left(\frac{1+is_{2n-1,2n}\gamma_{2n-1}^{s_{2n-1}}\gamma_{2n}^{s_{2n}}}{2}\right),
    \end{split}
    \label{eq:hprojS}
\end{align}

\begin{align}
    Y_c^{S,\{s^z,t^z\}}=2^{-\frac{n+1}{2}}(1+is_{2,3}\gamma_2^{s_2}\gamma_3^{s_3})(1+is_{4,5}\gamma_4^{s_4}\gamma_5^{s_5})\dots(1+is_{2n,1}\gamma_{2n}^{s_{2n}}\gamma_1^{s_1}),
    \label{eq:hfluctS}
\end{align}
where spins $s_i$ are determined by the pairing rule of Majorana fermions introduced in Sec.~\ref{subsec:wfn}. $s_{i,j}=1$ if the direction for $\langle v_iv_j\rangle$ is $\langle\overrightarrow{v_iv_j}\rangle$, and $s_{i,j}=-1$ for the opposite direction.
$C$ is Kasteleyn directed, since both $\mathrm{K}_i$ and $\mathrm{K}_f$ are Kasteleyn.

On the ground state Hilbert space of $H_{\mathrm{qubit}}+H_{\mathrm{decorate}}$, the operators given by the combination of~\eqref{eq:hprojS} and~\eqref{eq:hfluctS} associated with the bit flips, are shown to commute with each other, exactly in the same fashion as (2+1)d (see Sec.~V of~\cite{Tarantino}). This guarantees that the summand in $H_{\tau\mathrm{fluct}}+H_{\sigma\mathrm{fluct}}$ are commutative.

\section{Lattice model of (3+1)d $\mathbb{Z}_4^F$ trivial SPT}
\label{sec:3dz4f}
Based on a similar construction, we can also provide a (3+1)d model for the $\mathbb{Z}_4^F$ symmetric phase, with the Kitaev wire decorated on the wall.
$\mathbb{Z}_4^F$ denotes the unitary symmetry which squares to the fermion parity, $U^2=(-1)^F$.
We will construct the model on the (3+1)d lattice $\Gamma$, with the same degrees of freedom as Sec.~\ref{sec:3dt}. 
Both $\sigma$ and $\tau$ qubits are charged under the $\mathbb{Z}_4^F$ symmetry as the Pauli $x$,,
\begin{align}
    U_{\mathbb{Z}_4^F}: \ket{1}\mapsto\ket{0}, \ket{0}\mapsto\ket{1}.
    \label{eq:z4fqubit}
\end{align}
The edges of $\Gamma$ are also directed in the same way as Sec.~\ref{sec:3dt}. For Majorana fermions defined as~\eqref{eq:3dmajorana}, the $\mathbb{Z}_4^F$ symmetry acts as
\begin{align}
U_{\mathbb{Z}_4^F}:
\begin{cases}
a_v^{\uparrow}\to a_v^{\downarrow}\\
a_v^{\downarrow}\to-a_v^{\uparrow},
\end{cases}
\begin{cases}
b_{v'}^{\uparrow}\to b_{v'}^{\downarrow}\\
b_{v'}^{\downarrow}\to -b_{v'}^{\uparrow}.
\end{cases}
\end{align}
Then, we have the same $H_{\mathrm{qubit}}$ as Sec.~\ref{sec:3dham}, so that the domain wall of qubits form a 2d graph $\mathrm{K}$. Away from the Kitaev wire decoration, we pair up Majorana fermions along each short edge $\langle \overrightarrow{vv'}\rangle$, by $ia_v^{\uparrow}b_{v'}^{\uparrow}+ia_v^{\downarrow}b_{v'}^{\downarrow}$. 
The main difference from the $T$-SPT case is the way to pair up Majorana fermions on the domain wall. On the 1d domain wall of $\tau$ qubits on $\mathrm{K}$, we start with pairing up Majorana fermions along each long edge $\langle \overrightarrow{vw}\rangle$ bounding a planar triangle, by $i\gamma_v^{s_v}\gamma_w^{s_w}$.

Here, $s_v$ and $s_w$ are determined according to the following rule. We set up a local Cartesian coordinate system on the long edge.
The $x$ axis is set as a vector perpendicular to the 2d domain wall $\mathrm{K}$, directed from the domain of $\sigma$ qubits $\ket{1}$ to $\ket{0}$.
The $y$ axis is a vector parallel with $\mathrm{K}$, directed from the domain of $\tau$ qubits $\ket{1}$ to $\ket{0}$. Then, the $z$ axis can be defined as a vector parallel with the long edge, whose direction is fixed by the right-hand rule. Then,
\begin{itemize}
    \item If the $z$ axis is directed from $v$ to $w$ and the planar triangle is labeled by ``A'' (resp.~``B''), we have $s_v=\ \uparrow$ and $s_w=\ \downarrow$ (resp.~$s_v=\ \downarrow$ and $s_w=\ \uparrow$).
    \item If the $z$ axis is directed from $w$ to $v$ and the planar triangle is labeled by ``A'' (resp.~``B''), we have $s_v=\ \downarrow$ and $s_w=\ \uparrow$ (resp.~$s_v=\ \uparrow$ and $s_v=\ \downarrow$).
\end{itemize}
Here, the labels of planar triangles follows the rule in Fig.~\ref{fig:AB}.
After pairing up these Majorana fermions, we are left with one unpaired Majorana mode on each vertex of planar triangles, and two on each vertex of non-planar triangles. 
Then, we pair up yet unpaired Majorana fermions on short edges $\langle \overrightarrow{vw}\rangle$, as $i\gamma_v^{\overline{s}_v}\gamma_{w}^{\overline{s}_w}$. 
Here, we will choose the pairing such that $\overline{s}_v=\overline{s}_w$.
Finally, we have one unpaired Majorana fermion on each vertex of non-planar triangles. We pair them up along long edges of non-planar triangles $\langle \overrightarrow{vw}\rangle$ as $i\gamma_v^{s_v}\gamma_{w}^{s_w}$. Here, we can see that $s_v$ is the same as $s_w$, $s_v=s_w$.

The wave function of the $\mathbb{Z}_4^F$ SPT phase is given by the equal superposition of all possible configurations of $\sigma$ qubits and $\tau$ qubits. To see the invariance of the wave function under the $\mathbb{Z}_4^F$ symmetry, we first note that the pairing of Majorana fermions away from the Kitaev wire decoration $ia_v^{\uparrow}b_{v'}^{\uparrow}+ia_v^{\downarrow}b_{v'}^{\downarrow}$ is invariant under $U_{\mathbb{Z}_4^F}$. 
On the Kitaev wire decoration, according to the pairing rule, the spins of paired Majorana fermions $s_v, s_w$ flip their signs under the action of $U_{\mathbb{Z}_4^F}$, which is consistent with the transformation law of fermions. This is because the labels of planar triangles (``A'' or ``B'') are changed under $U_{\mathbb{Z}_4^F}$, and the direction of the local $z$ axis on the long edges bounding planar triangles is invariant, thereby the spins of paired Majorana fermions are flipped. 

We can also check that the pairings of Majorana fermions are consistent with the induced Kasteleyn direction on $\mathrm{K}$ under the $U_{\mathbb{Z}_4^F}$ transformation. 
On one hand, on short edges and long edges bounding non-planar triangles of $\mathrm{K}$, the sign of the pairing $i\gamma_v^{s_v}\gamma_{w}^{s_w}$ is invariant under $U_{\mathbb{Z}_4^F}$, which is consistent with the invariance of the direction on $\langle vw\rangle$. 
On the other hand, on long edges bounding planar triangles of $\mathrm{K}$, the pairing $i\gamma_v^{s_v}\gamma_{w}^{s_w}$ flips its sign under $U_{\mathbb{Z}_4^F}$. 
It is also consistent with the Kasteleyn direction on $\mathrm{K}$, which flips the directions.

The Hamiltonian is defined in exactly the same form as Sec.~\ref{sec:3dham}, except for the pairing rule of Majorana fermions (i.e., the choice of spins of Majorana fermions in the expressions~\eqref{eq:hdeco},~\eqref{eq:pif} and~\eqref{eq:xf}), so we will not repeat the construction here.

\section{Analysis of the SPT phase}
\label{sec:wallspt}

\subsection{time-reversal case}
Here, we claim that our $T$-SPT model constructed in Sec.~\ref{sec:3dt} generates the $\mathbb{Z}_8$ subgroup in the $\mathbb{Z}_{16}$ classification. 
To see this, it is convenient to study the time reversal domain wall of the model.
Suppose we have prepared a topological quantum field theory (TQFT) that describes our model at long distances. The $T$ symmetry with $T^2=(-1)^F$ means that we can place the TQFT on an (3+1)d unoriented spacetime manifold $X$ equipped with a pin$^+$ structure. By breaking the $T$ symmetry, we obtain a codimension-1 worldvolume of $T$ domain wall $Y$ between domains with the broken symmetry, which can support a nontrivial topological theory. 

Then, let us examine the spacetime structure induced on $Y$, which amounts to identifying the symmetry of the domain wall. We note that $Y$ does not admit pin$^+$ in general, since the normal bundle $NY$ can twist the structure in a nontrivial fashion~\cite{Kapustin:2014dxa}. 
To see the structure of $Y$, we express the pin$^+$ structure in terms of the spin structure on $TX\oplus\rho$, with $\rho=L^{\mathrm{or}}\oplus L^{\mathrm{or}}\oplus L^{\mathrm{or}}$ for the orientation line bundle $L^{\mathrm{or}}$. 
We further prepare $L^{\mathrm{or}}$ in terms of the pull-back of the universal line bundle $\sigma$ on $B\mathbb{Z}_2$ given by the sign representation, by a $\mathbb{Z}_2$ gauge field $A: X\mapsto B\mathbb{Z}_2$; $L^{\mathrm{or}}=A^*\sigma$. 
Then, $Y$ is defined as the zero locus of the section of $L^{\mathrm{or}}=A^*\sigma$.
Since we have $NY=A^*\sigma$ when restricted to $Y$, the induced structure on $Y$ becomes the spin structure on
\begin{align}
    TY\oplus NY\oplus 3A^*\sigma = TY\oplus 4A^*\sigma,
    \label{eq:Ystructure}
\end{align}
which is equivalent to having a spin structure on $TY$ with a $\mathbb{Z}_2$ gauge field $A$. Hence, the theory on the domain wall has the unitary $\mathbb{Z}_2$ symmetry, whose gauge field is given by restricting the orientation line bundle to the domain wall.

Meanwhile, in our model we prepare a codimension-1 domain wall by breaking the $T$ symmetry; we set a spatial manifold as $M=N\times\mathbb{R}$ for some 2d manifold $N$, and turn on the ferromagnetic Ising interaction $\sigma^z_c\sigma^z_{c'}$ for $\sigma$ qubits. 
By introducing frustrated boundary conditions fixing $\sigma$ qubits at $x\to\infty$ (resp.~$x\to-\infty$) as $\ket{1}$ (resp.~$\ket{0}$), we get a 2d domain wall of $\sigma$ qubits between the domains with the spontaneously broken symmetry. Once we fix the configuration of the domain wall, we have fluctuating $\tau$ qubits on a 2d graph $\mathrm{K}$ of the domain wall. 
Since both $\sigma$ and $\tau$ qubits are charged under time reversal, we can identify these qubits as a placeholder for the section of (time slice of) the orientation bundle $A$. Therefore, the $\tau$ qubit works as a placeholder for the section of the induced $\mathbb{Z}_2$ gauge field on the domain wall, indicating the presence of the unitary $\mathbb{Z}_2$ symmetry on the domain wall, which charges the $\tau$ qubits. 

We actually find such a $\mathbb{Z}_2$ symmetry acting on the domain wall; the effective Hamiltonian for the fixed 2d domain wall in the symmetry broken phase is given by
\begin{align}
    H=H_{\mathrm{decorate}}|_{K}+H_{\tau\mathrm{fluct}}|_{K},
    \label{eq:2deff}
\end{align}
where $H_{\mathrm{decorate}}|_{K}$ realizes the effective action of $H_{\mathrm{decorate}}$ on $\mathrm{K}$ in~\eqref{eq:hdeco},
\begin{align}
    H_{\mathrm{decorate}}|_{K}=\sum_{\substack{\mathrm{short\ edge}\\ \langle\overrightarrow{vw}\rangle}}(ia_v^{\uparrow}b_{w}^{\uparrow}+ia_v^{\downarrow}b_{w}^{\downarrow})(1-D_{\langle vw\rangle})+\sum_{\langle \overrightarrow{vw}\rangle}(i\gamma_v^{s_v}\gamma_{w}^{s_w})D_{\langle vw\rangle},
\end{align}
for $D_{vw}= (1-\tau_f^z\tau_{f'}^z)/2$, where $f, f'$ are faces of $\mathrm{K}$ sandwiching the edge $\langle vw\rangle$.~\footnote{Here, we locate a fictitious  $\tau$ qubit on each triangle on $\mathrm{K}$ according to the majority rule, as we have done in the (2+1)d case in Sec.~\ref{sec:tarantino}.} We also have the effective action for $H_{\tau\mathrm{fluct}}$ in~\eqref{eq:htfluct},
\begin{align}
    H_{\tau\mathrm{fluct}}|_{K}=\sum_f \tau_f^x X_f,
\end{align}
where $X_f$ rearranges the Majorana pairings inside $\mathrm{K}$,
\begin{align}
    X_f=\sum_{\{t^z\};  R_f}X_f^{\{t^z\}}\Pi_f^{\{t^z\}}P_f^{\{t^z\}},
\end{align}
with $P_f^{\{t^z\}}$ defined as~\eqref{eq:pf}, and $\Pi_f^{\{t^z\}}$, $P_f^{\{t^z\}}$ defined as~\eqref{eq:pif},~\eqref{eq:xf} respectively.
Though we defined the directions of edges on $\mathrm{K}$ which are sometimes reversed by time reversal, we obtain a fixed Kasteleyn direction on $\mathrm{K}$ after breaking time reversal.
The (2+1)d model~\eqref{eq:2deff} essentially realizes the nontrivial $\mathbb{Z}_2$ SPT phase introduced in Sec.~\ref{sec:tarantino}, based on the $\mathbb{Z}_2$ symmetry defined as 
\begin{align}
    U_{\mathbb{Z}_2}=\prod_f \tau_f^x,
\end{align}
and leaves the fermions invariant. Thus, upon breaking the $T$ symmetry we have the unbroken $\mathbb{Z}_2$ symmetry on the 2d domain wall, which protects the Tarantino-Fidkowski SPT phase.
In general, there is of course an ambiguity regarding the definition of the $\mathbb{Z}_2$ symmetry, e.g., we could redefine $U_{\mathbb{Z}_2}$ by combining with another internal symmetries on the wall, such as fermion parity. In a relativistic theory, the canonical way to obtain the induced symmetry is believed to exist with the help of the unbreakable CPT symmetry~\cite{HKT2019CPT}.  
Though the discrete lattice analog without Lorentz invariance for such a mechanism is left unspecified, we have the SPT phase with an odd $\mathbb{Z}_8$ index regardless of such an ambiguity in our case, once we claim that the $\tau$ qubits must be charged under $U_{\mathbb{Z}_2}$.

Such a study on the symmetry of the domain wall is quite useful to find the classification. $(d+1)$d $T$-SPT phases with $T^2=(-1)^F$ are classified by the cobordism group $\Omega^{d+1}_{\mathrm{spin}}(B\mathbb{Z}_2, 3\sigma)$, which is the Anderson dual of the bordism group $\Omega_{d+1}^{\mathrm{spin}}(B\mathbb{Z}_2, 3\sigma)$. Here, $\Omega^{d+1}_{S}(BG, \xi)$ with $\xi$ a real vector bundle over $BG$, is the bordism group of the triple $[X, A, s]$, where $X$ is a $(d+1)$d manifold, $A: X\mapsto BG$ is a $G$-bundle over $X$, and $s$ is a $S$-structure on $TX\oplus A^*\xi$.

The restriction to the zero locus $Y$ of the section of $A^*\sigma$ induces the map between bordism groups,
\begin{align}
    \Omega_{d+1}^{\mathrm{spin}}(B\mathbb{Z}_2, 3\sigma)\mapsto \Omega_{d}^{\mathrm{spin}}(B\mathbb{Z}_2, 4\sigma),
\end{align}
since $Y$ admits the spin structure on $TY\oplus 4A^*\sigma$, as we have seen in~\eqref{eq:Ystructure}.
Dually, we also have the map between cobordism groups. In~\cite{HKT2019CPT}, it has been shown that the dual map
\begin{align}
    \Omega^{d}_{\mathrm{spin}}(B\mathbb{Z}_2, (m+1)\sigma)\mapsto \tilde{\Omega}^{d+1}_{\mathrm{spin}}(B\mathbb{Z}_2, m\sigma)
    \label{eq:smith}
\end{align}
is always surjective for $m\in\mathbb{Z}$, where $\tilde{\Omega}^{d+1}_{S}(BG, \xi)$ is the reduced cobordism group, dual to the reduced bordism group $\tilde{\Omega}_{d+1}^{S}(BG, \xi)$. $\tilde{\Omega}_{d+1}^{S}(BG, \xi)$ is defined as the cokernel of the map $\Omega_{d+1}^{S}\mapsto\Omega_{d+1}^{S}(BG, \xi)$, given by equipping with the trivial $G$ gauge field.
In particular, for $m=3, d=3$ the above map is $\mathbb{Z}_8\oplus\mathbb{Z}\mapsto\mathbb{Z}_{16}$, which is determined as~\cite{HKT2019CPT}
\begin{align}
    (\nu, k)\mapsto 2\nu-k \quad \mod 16.
\end{align}
This allows us to obtain the classification of (3+1)d $T$-SPT phases by the (2+1)d SPT phases with the induced symmetry on the domain wall. The Tarantino-Fidkowski $\mathbb{Z}_2$ SPT phase corresponds to odd $\nu$ in $\mathbb{Z}_8$ with $k=0$ (i.e., no gravitational anomaly on the boundary). Hence, our model realizes $2\nu\in \mathbb{Z}_{16}$ for odd $\nu$, which generates the $\mathbb{Z}_8$ subgroup in the full classification.

\subsection{$\mathbb{Z}_4^F$ case}
A similar argument also applies for the $\mathbb{Z}_4^F$ symmetric model in Sec.~\ref{sec:3dz4f}. $\mathbb{Z}_4^F$ symmetry corresponds to the $\mathrm{spin}^{\mathbb{Z}_4}:=(\mathrm{spin}\times\mathbb{Z}_4)/\mathbb{Z}_2$ structure, which is equivalently the spin structure on $TX\oplus 2A^*\sigma$. Thus, the induced structure on the worldvolume of the $U_{\mathbb{Z}_4^F}$ domain wall $Y$ becomes the spin on
\begin{align}
    TY\oplus NY\oplus 2A^*\sigma = TY\oplus 3A^*\sigma,
    \label{eq:Ystructure23}
\end{align}
which is equivalent to the pin$^+$ structure on $Y$. Hence, on the 2d domain wall of the $U_{\mathbb{Z}_4^F}$ symmetry, we expect an anti-unitary symmetry $T$ which squares to the fermion parity $T^2=(-1)^F$, charging the $\tau$ qubits. Once we fix the configuration of the 2d domain wall $\mathrm{K}$, we can actually find such a $T$ symmetry, as we will see in Appendix~\ref{app:z4f}.
Under this $T$ symmetry, the domain wall hosts a nontrivial (2+1)d $T$-SPT phase equivalent to the model constructed in~\cite{Zitao}. 

The relevant map of cobordism groups in our case is $m=2, d=3$ in~\eqref{eq:smith}, which gives a zero map $\mathbb{Z}_2\mapsto 0$. A puzzle here is that the classification of the (3+1)d $\mathbb{Z}_4^F$ SPT phase is trivial, $\Omega_{\mathrm{spin}}^4(B\mathbb{Z}_2, 2\sigma)=0$~\cite{Garcia-Etxebarria:2018}. So, somehow there should be a $\mathbb{Z}_4^F$ symmetric deformation that transforms the model into a trivial atomic insulator. This point will be considered in future work.

\section*{Acknowledgements}
The author is grateful to Yu-An Chen and Ryan Thorngren for useful discussions.
The author also acknowledges the hospitality of Harvard CMSA.
The author is supported by Japan Society for the Promotion of Science (JSPS) through Grant No.~19J20801. 

\appendix
\section{A formula for Stiefel-Whitney homology classes}
\label{app:w2}
In this appendix, we prove the expression for the dual of the representative of $w_2$~\eqref{eq:w2formula}. First we recall the theorem in~\cite{Chen2019bosonization},
\begin{description}
\item[Theorem.] In a 3d manifold $M$ with triangulation  and branching structure, the homology class of the dual of $w_2$ is represented by a 1-chain $S'\in C_1(M, \mathbb{Z}_2)$, 
\begin{align}
    S'= \sum_{e}e-\sum_{\Delta_+=\langle0123\rangle} \langle02\rangle-\sum_{\Delta_-=\langle0123\rangle} \langle13\rangle,
    \label{eq:w2chen}
\end{align}
where the first sum is over all 1-simplices of the triangulation, and $\Delta_+$ (resp.~$\Delta_-$) denotes a $+$ (resp.~$-$) 3-simplex.
\end{description}
We show that the above 1-chain $S'$ is homologically equivalent to $S$ in~\eqref{eq:w2formula}. 
To do this, we consider a branching structure of $\mathcal{T}'$ defined as follows. 
First, we assign a local ordering to vertices of $\mathcal{T}$, such that the vertex on the barycenter of a $i$-simplex is labeled as $i$.
Then, while respecting the ordering on vertices of $\mathcal{T}$, we further assign a local ordering on vertices of $\mathcal{T}'$, such that a barycenter of a $j$-simplex of $\mathcal{T}$ has a larger ordering than that of an $i$-simplex if $j>i$. Then, we have an induced branching structure on $\mathcal{T}'$.
Based on this branching structure, after some efforts we can write $S'$ in~\eqref{eq:w2chen} as
\begin{align}
\begin{split}
    S'=&\sum_{e\in\mathcal{T}'}e-\sum_{\Delta_+}(\langle v_1v_{0123}\rangle+\langle v_3v_{0123}\rangle+\langle v_1v_{012}\rangle+\langle v_2v_{023}\rangle)
    \\
    &-\sum_{\Delta_-}(\langle v_1v_{0123}\rangle+\langle v_3v_{0123}\rangle+\langle v_1v_{013}\rangle+\langle v_2v_{123}\rangle),
\end{split}
\end{align}
where the convention is the same as the expression in~\eqref{eq:w2formula}.
Up to a boundary of a 2-chain, the above $S'$ is written as
\begin{align}
    S'=\sum_{e\in\mathcal{T}'}e-\sum_{\Delta_+}(\langle v_{012}v_{0123}\rangle+\langle v_{023}v_{0123}\rangle+\langle v_2v_3\rangle)-\sum_{\Delta_-}(\langle v_{013}v_{0123}\rangle+\langle v_{123}v_{0123}\rangle+\langle v_2v_3\rangle).
\end{align}
Since the contributions of $\langle v_2v_3\rangle$ cancel out on $+$ and $-$ simplices, we finally get~\eqref{eq:w2formula}.

\section{Detailed descriptions of $H_{\mathrm{qubit}}$}
\label{app:hqubit}

\begin{itemize}
    \item Each $\sigma$ qubit on the barycenter of a 2-simplex $\langle v_0v_1v_2\rangle$ of $\mathcal{T}$ is fixed according to the majority rule.
    This is done by introducing the term,
    \begin{align}
        H_{012}=-\sum_{\langle v_0v_1v_2\rangle}\sum_{\sigma_v^z}P_{012}(\sigma^z_{012}; \sigma^z_{0}, \sigma^z_{1}, \sigma^z_{2}),
    \end{align}
    where the second sum runs over $2^3=8$ types of configuration of three $\sigma$ qubits $\sigma^z_0, \sigma^z_1, \sigma^z_2$ on vertices $v_0,v_1,v_2$ respectively. $P_{012}(\sigma^z_{012}; \sigma^z_{0}, \sigma^z_{1}, \sigma^z_{2})$ denotes the projector which stabilizes the majority of three qubits for the $\sigma$ qubit $\sigma^z_{012}$ at $v_{012}$. 
    For instance, if $\sigma^z_{0}=\sigma^z_{1}=\sigma^z_{2}=1$, $P_{012}(\sigma^z_{012}; \sigma^z_{0}, \sigma^z_{1}, \sigma^z_{2})$ acts on $\sigma^z_{012}$ as $(1+\sigma^z_{012})/2$. 
    
    \item Each $\sigma$ qubit on the barycenter of a 3-simplex $\langle v_0v_1v_2v_3\rangle$ of $\mathcal{T}$ is also determined by the majority rule. 
    This is done by introducing the term,
    \begin{align}
        H_{0123}=-\sum_{\langle v_0v_1v_2v_3\rangle}\sum_{\sigma_v^z}P_{0123}(\sigma^z_{0123}; \sigma^z_{0}, \sigma^z_{1}, \sigma^z_{2}, \sigma^z_{3}),
    \end{align}
    where the second sum runs over $2^4=16$ types of configuration of four $\sigma$ qubits on vertices $v_0,v_1,v_2, v_3$. $P_{0123}(\sigma^z_{012}; \sigma^z_{0}, \sigma^z_{1}, \sigma^z_{2}, \sigma^z_{3})$ denotes the projector which stabilizes the majority of four qubits for the $\sigma$ qubit $\sigma^z_{0123}$ at $v_{0123}$. 
    For instance, if $\sigma^z_{0}=\sigma^z_{1}=\sigma^z_{2}=1$ and  $\sigma^z_3=-1$, $P_{0123}(\sigma^z_{0123}; \sigma^z_{0}, \sigma^z_{1}, \sigma^z_{2}, \sigma^z_{3})$ acts on $\sigma^z_{0123}$ as $(1+\sigma^z_{0123})/2$.
    If exactly two of four qubits at $v_0,v_1,v_2, v_3$ have $\ket{1}$, we take $P_{0123}=\mathrm{id}$.
    
    \item Each $\tau$ qubit on the face of $\Gamma$ is fixed if it is away from $\mathrm{K}$, depending on the domain of $\sigma$ qubits: $\ket{1}$ (resp.~$\ket{0}$) if contained in the domain of $\ket{1}$ (resp.~$\ket{0}$).
    This is done by introducing the term,
    \begin{align}
        H_{\mathrm{away}\tau}=-\sum_f\left(\frac{1+\tau_f^z\sigma_{c'}^z}{2}\cdot \frac{1+\sigma_c^z\sigma_{c'}^z}{2}+\frac{1-\sigma_c^z\sigma_{c'}^z}{2}\right),
    \end{align}
    where $f$ denotes the face of $\Gamma$, and $c$, $c'$ are two 3-cells of $\Gamma$ sandwiching $f$. The sum runs over all $\tau$ qubits on $\Gamma$.
    
    \item The 2d graph $\mathrm{K}$ was obtained by gathering four faces in $\mathrm{K}'$ into a single face, which was described in Fig.~\ref{fig:kpk}. 
    Since we have one $\tau$ qubit on each face of $\mathrm{K}'$ except for triangles, the newly obtained single face of $\mathrm{K}$ in Fig.~\ref{fig:kpk} contains two $\tau$ qubits.
    These two $\tau$ qubits on faces $f, f'$ of $\mathrm{K}'$ share the same state, i.e., $\ket{00}$ or $\ket{11}$. It is realized by the term,
    \begin{align}
        H_{\mathrm{dw}\tau}=\sum_{\langle v_0v_1v_2v_3\rangle}\sum_{\sigma_v^z}P_{\tau}(\tau^z_f, \tau^z_{f'}; \sigma_0^z, \sigma_1^z, \sigma_2^z, \sigma_3^z, \sigma_{0123}^z).
    \end{align}
    Here, $P_{\tau}$ is an operator which acts as $(1-\tau_f^z\tau_{f'}^z)/2$, when exactly two of four $\sigma$ qubits on $v_0,v_1,v_2,v_3$ are $\ket{1}$, where the configuration of  $\mathrm{K}'$ looks like Fig.~\ref{fig:2dgraph} (b). 
    Otherwise, $P_{\tau}$ acts as zero. The second sum runs over $2^5=32$ patterns of five qubits $\sigma_0^z, \sigma_1^z, \sigma_2^z, \sigma_3^z, \sigma_{0123}^z$, which determines the configuration of $\mathrm{K}'$ on a 3-simplex $\langle v_0v_1v_2v_3\rangle$.
\end{itemize}

Summarizing, $H_{\mathrm{qubit}}$ is written as
\begin{align}
    H_{\mathrm{qubit}}=H_{012}+H_{0123}+H_{\mathrm{away}\tau}+H_{\mathrm{dw}\tau}.
\end{align}

\section{Symmetry of the domain wall for the (3+1)d $\mathbb{Z}_4^F$ SPT}
\label{app:z4f}
Here, we study the induced symmetry on the 2d domain wall of the (3+1)d $\mathbb{Z}_4^F$ symmetric phase in Sec.~\ref{sec:3dz4f}. We fix the configuration of the (2+1)d domain wall $\mathrm{K}$, by freezing $\sigma$ qubits via spontaneous breaking of the $\mathbb{Z}_4^F$ symmetry. The wave function on $\mathrm{K}$ is given by decorating the Kitaev wire on the fluctuating domain wall of $\tau$ qubits, which is again a ground state of the effective Hamiltonian in the form of~\eqref{eq:2deff}. The wave function has the following $T$ symmetry; $T$ acts on $\tau$ qubits as the Pauli $x$ operator
\begin{align}
    T: \ket{1}\mapsto\ket{0}, \ket{0}\mapsto\ket{1}.
\end{align}
Then, we can define the symmetry action on Majorana fermions on $\mathrm{K}$ with $T^2=(-1)^F$ in the following way. First, we label each Majorana fermion on $\mathrm{K}$ by ``$\alpha$'' or ``$\beta$'', such that 
\begin{itemize}
    \item for long edges $\langle vw\rangle$ bounding a planar triangle, Majorana fermions $\gamma_v$ and $\gamma_w$ share the same label.
    \item for other edges $\langle vw\rangle$ bounding $\mathrm{K}$, $\gamma_v$ and $\gamma_w$ have different labels.
\end{itemize}
For instance, we can label Majorana fermions on vertices of ``A'' triangles by ``$\alpha$'', and vertices of ``B'' triangles by ``$\beta$''. (Here, the labels of planar triangles follows Fig.~\ref{fig:AB}.) The rest of fermions are automatically labeled by requiring the above rule. 

Then, we define the symmetry action on $\mathrm{K}$ as
\begin{align}
T:
\begin{cases}
\alpha_v^{\uparrow}\to \alpha_v^{\downarrow}\\
\alpha_v^{\downarrow}\to -\alpha_v^{\uparrow},
\end{cases}
\begin{cases}
\beta_{v'}^{\uparrow}\to -\beta_{v'}^{\downarrow}\\
\beta_{v'}^{\downarrow}\to \beta_{v'}^{\uparrow}.
\end{cases}
\end{align}
We can see the invariance of the wave function on $\mathrm{K}$ under $T$ as follows. First, we are pairing the Majorana fermion away from the 1d domain wall, along the short edges $\langle \overrightarrow{vv'}\rangle$ like $\pm(i\alpha_v^{\uparrow}\beta_{v'}^{\uparrow}+i\alpha_v^{\downarrow}\beta_{v'}^{\downarrow})$, which is invariant under the symmetry. Next, let us examine the 1d domain wall.
Since the $T$ flips $\tau$ qubits on $\mathrm{K}$, we flip the local $y$ axis along the edges on the domain wall, which is explained in Sec.~\ref{sec:3dz4f}. Accordingly, $T$ flips the $z$ axis according to the right hand rule, thereby flipping the spins of fermions on the domain wall. This is consistent with the required symmetry action of $T$ on Majorana fermions.
We can also show the consistency of the pairing of Majorana fermions with the Kasteleyn orientation under the $T$ action.

The wave function on $\mathrm{K}$ is essentially the same as the model introduced in~\cite{Zitao}, which realizes the (2+1)d $T$-SPT phase.

\bibliographystyle{ytphys}
\baselineskip=.95\baselineskip
\bibliography{ref}

\end{document}